\documentclass[11pt,a4paper]{article}

\usepackage{a4wide,lmodern,graphicx,verbatim,authblk}
\usepackage[colorlinks=true,hyperfootnotes=false,citecolor=blue]{hyperref}
\usepackage{amsmath,amssymb}

\newtheorem{rmrk}{Remark}

\newcommand{\x}{\mathbf{x}}

\newcommand{\f}{\mathbf{f}}
\newcommand{\q}{\mathbf{q}}
\newcommand{\tf}{\tilde{\f}}

\newcommand{\U}{\mathbf{u}}

\newcommand{\hU}{\hat{\U}}
\newcommand{\tU}{\tilde{\U}}
\newcommand{\Uself}{\mathbf{u}_{\text{self}}}
\newcommand{\id}{\mathbf{I}}

\newcommand{\Ewp}{\mathrm{Ew}_\mathrm{2P}}

\newcommand{\quadcorr}{\Upsilon_\mathrm{loc}}

\newcommand{\y}{\mathbf{y}}

\renewcommand{\Re}{\mathrm{Re}\,}
\renewcommand{\Im}{\mathrm{Im}\,}

\newcommand{\pd}[2]{\frac{\partial #1}{\partial #2}}

\renewcommand{\k}{\mathbf{k}}
\newcommand{\nk}{\|\k\|}
\newcommand{\kf}{\Bbbk}
\newcommand{\nkf}{\|\Bbbk\|}
\newcommand{\NL}{\mathrm{NL}}
\newcommand{\rms}{\mathrm{rms}}

\newcommand{\Ufz}{\U^{F,\k=0}}

\newcommand{\p}{\mathbf{p}}

\newcommand{\expk}[1]{e^{-i\k\cdot #1}}
\renewcommand{\d}{\mathrm{d}}
\renewcommand{\r}{\rho}

\newcommand{\R}{\mathbb{R}}
\newcommand{\Z}{\mathbb{Z}}
\newcommand{\erf}{\text{erf}}
\newcommand{\erfc}{\text{erfc}}
\renewcommand{\O}{\mathcal{O}}
\newcommand{\kmax}{k_\infty}
\newcommand{\epsi}{\varepsilon}
\newcommand{\sumn}{\sum_{n=1}^N}
\newcommand{\sump}[1]{\sum_{\p\in \Z^#1}}
\newcommand{\tp}{\tau(\p)}
\newcommand{\tQ}{\tilde{Q}}

\newcommand{\QI}{Q^\id}
\newcommand{\Qkk}{Q^{\kf\otimes\kf}}

\title{Fast and spectrally accurate summation of 2-periodic Stokes potentials}
\author[1,*]{Dag Lindbo}
\author[1]{Anna-Karin Tornberg} 
\affil[1]{Numerical Analysis, Royal Inst. of Tech. (KTH), 100 44
  Stockholm, Sweden}

\begin{document}
\maketitle

\thispagestyle{empty}
\let\oldthefootnote\thefootnote
\renewcommand{\thefootnote}{\fnsymbol{footnote}}
\footnotetext[1]{To whom correspondence should be addressed. Email:
  \url{dag@kth.se} }
\let\thefootnote\oldthefootnote

\begin{abstract}
  We derive a Ewald decomposition for the Stokeslet in planar
  periodicity and a novel PME-type $\O(N \log N)$ method for the fast
  evaluation of the resulting sums. The decomposition is the natural
  2P counterpart to the classical 3P decomposition by Hasimoto, and is
  given in an explicit form not found in the literature. Truncation
  error estimates are provided to aid in selecting parameters. The
  fast, PME-type, method appears to be the first fast method for
  computing Stokeslet Ewald sums in planar periodicity, and has three
  attractive properties: it is spectrally accurate; it uses the
  minimal amount of memory that a gridded Ewald method can use; and
  provides clarity regarding numerical errors and how to choose
  parameters. Analytical and numerical results are give to support
  this. We explore the practicalities of the proposed method, and
  survey the computational issues involved in applying it to
  2-periodic boundary integral Stokes problems.
\end{abstract}

\section{Introduction}

Viscous flow systems continuously enjoy much attention from various
scientific disciplines, such as the widespread study of passive and
motile suspensions and other applications in \mbox{bio-,} micro- and complex
fluidics. An example is the large body of work that involves the study
of locomotion of small organisms, the basics of which are illustrated
in a classic article by Purcell \cite{Purcell1977}, with further
developments in e.g. Koiller et al. \cite{Koiller1996}. Theoretical
understanding of various modes of propulsion \cite{Spagnolie2010,
  Spagnolie2010a, GonzalezRodriguez2009, Childress2011, Ishikawa2006,
  Ishikawa2007} is rapidly progressing apace with simulation methods
\cite{Kanevsky2010, Saintillan2007}. If history is any guide,
analytical results will feed into computational models and rich
computational studies of complex systems will emerge. There are also
notably strong interdisciplinary connections present in this area. A
recent survey of the modelling of biomimetic fluid flow by Saha
et al. \cite{Saha2011} provides a broader perspective, including
modeling and computation outside of the viscous flow regime.

Another area where there has been much activity is in simulation of
suspensions of various particles in viscous flows, motivated, for
instance, by a desire to understand the formation of microstructures
\cite{Hosein2007} and paper-making \cite{Lundell2011}. Here one finds
detailed computational studies of rigid or flexible fiber suspension,
such as the work by Saintillan et al. \cite{Saintillan2005} and
Tornberg, Shelley and collaborators \cite{Tornberg2004, Tornberg2006,
  Shelley2000}, as well as large body of work concerning suspensions
of various spheroids \cite{Butler2002,Sierou2001} and related
particles \cite{Kumar2011}. Complex structure formation is observed,
based on accurate modeling of hydrodynamical interactions, in these
computational investigations.

Alas, the present work shall not in itself elucidate these fascinating
questions of collective dynamics, but rather pursue algorithmic
development that could help such investigations deal with larger
systems in future.

The models under consideration in the works cited, and in countless
other investigations, are based on the Stokes equations. A variety of
numerical and analytical techniques are used, but a common feature is
the use of singularity solutions. One class of methods is the so
called distributed singularity approach, in which the various Green's
functions of Stokes equations are combined in point, line, or surface
distributions to generate the flow induced by a particular particle or
distribution. In this way, slender bodies (e.g. fibers) are
represented by a line distribution of Stokes potentials along its
centerline, as was suggested by Batchelor \cite{Batchelor1970}. In an
analogous way, spheroids can be represented by a combination of
higher-order Green's functions, see e.g. Zhou \& Pozrikidis
\cite{Zhou1995} and G\"{o}tz \cite{Gotz2005}.

There are also direct boundary integral methods, as discussed in
e.g. Pozrikids \cite{Pozrikidis2001, Pozrikidis1992}, Anderson
et al. \cite{Janssen2007, Bazhlekov2006, Bazhlekov2004}, and others
\cite{Keaveny2011,Ying2006,Young2009}. Another class of methods that
has been highly successful is known as \emph{Stokesian dynamics}, and
is due to Brady and collaborators \cite{Brady1988, Sierou2001}. All of
these methods are based on distributions of Green's functions that are
either summed or integrated, often in intricate ways.

Moreover, results free from finite-size effects are often of interest;
the well-trodden path towards which is applying periodic boundary
conditions to a sufficiently large system \cite{Sierou2001,
  Tornberg2004, Saintillan2005}. Solvers that exploit periodic
structure, such as the fast Ewald methods that we shall survey
shortly, are often highly specialized. In particular, there are
fundamental differences between how full (i.e. applied in all three
directions) and lower dimensional periodicity (i.e. periodicity with
respect to one or two dimensions) enters in mathematical
(cf. Pozrikidis \cite{Pozrikidis1996}) and algorithmic terms. Whereas
methods are relatively well established for fully periodic Stokes
problems \cite{Sierou2001,Saintillan2005,Lindbo2010}, methods for
problems in planar periodicity, e.g. when periodicity is applied to
$(x,y)$ and $z$ is ``free'', are less so (see
e.g. \cite{Pozrikidis1996,Ishii1979}). However, such systems,
including wall-confined systems, applications in biology (such as
beating flagella in planar configurations
\cite{Coq2011,Khaderi2009,Kokot2011}), are of growing interest.

The present work deals with the efficient, $\O(N \log N)$, and
spectrally accurate summation of Stokes potentials in the 2P
setting. To clarify, let $\Omega = [0, L)^3$ and consider Stokes
equations for $\x\in\Omega$,
\begin{align}
  \begin{split}
    -\nabla p(\x) + \mu \Delta \U(\x) = \f(\x)\\
    \nabla \cdot \U(\x) = 0,
  \end{split} \label{eq:stokes_eq}
\end{align}
where $\U(\x)$ denotes the velocity field, $p(\x)$ the pressure, and
$\f(\x)$ the force applied to the fluid. The fundamental solution of
Stokes flow represents solutions of the singularly forced Stokes
equations, i.e. \eqref{eq:stokes_eq} with
$\f(\x)=\f_0\delta(\x-\x_0)$.

Introducing the \emph{Stokeslet},
\begin{align}
  S(\x) = \frac{\id}{\|\x\|} + \frac{\x \otimes
    \x}{\|\x\|^3}, \label{eq:stokeslet}
\end{align}
where $(\id)_{ij} = \delta_{ij}$ is the identity and $\x\otimes\x$ is
the outer product $x_i x_j$, the solution of \eqref{eq:stokes_eq} in
free space with $\f(\x)=\f_0\delta(\x-\x_0)$ can be written:
\begin{align*}
  \U(\x) = \frac{1}{8\pi\mu}\int_{\R^3} S(\x-\y) \f_0 \delta(\y -
  \x_0 ) \d\y = \frac{1}{8\pi\mu} S(\x-\x_0) \f_0. 
\end{align*}
See e.g. the textbook by Pozrikidis \cite[p. 22]{Pozrikidis1992} for a
derivation of \eqref{eq:stokeslet}. A solution $\U(\x), \x\in\Omega$,
that satisfies periodicity is expressed as an infinite sum over
periodic images of $\f_0$ convolved with $S$, the Stokeslet,
\begin{align*}
  \U(\x) &= \frac{1}{8\pi\mu}\int_{\R^3} S(\x-\y) \sump{d} \f_0
  \delta(\y - \x_n + \tp) \d\y \\
  &= \frac{1}{8\pi\mu}\sump{d} S(\x - \x_n +
  \tp)\f_0,
\end{align*}
where $d=1,2,3$ is the dimension of periodicity (recall, $\x\in\R^3$
throughout) and $\tau:\Z^d \rightarrow \R^3$ is a translation into the
periodic image $\p$. Under fully periodic boundary conditions (3P) one
would let $\tp=L\p$, and in the 2-periodic (2P) case we let $\tp=[L\p,
0]$.

In light of the applications surveyed above, we let $\f$ denote an
array of point forces,
\begin{align}
  \f(\x) = \sumn \f_n \delta(\x-\x_n), \label{eq:f_deltas}
\end{align}
where $\f_n$ may contain any particular set of e.g. quadrature weights
and physical constants. We focus on the evaluation the periodized
Stokeslet sum
\begin{align}
  \U(\x)=\sump{d} \sumn S(\x - \x_n +
  \tp)\f_n, \label{eq:stokeslet_sum}.
\end{align}

Note that the terms in \eqref{eq:stokeslet_sum} decay as $1/r$, and
\eqref{eq:stokeslet_sum} is obviously not absolutely
summable. Takemoto et al. \cite{Takemoto2003} discuss this in the
related context of the electrostatic potential. However, the practical
computation of \eqref{eq:stokeslet_sum} is plainly infeasible, even
for small $N$, so fast methods are essential. Alternatives exist at
this point, including pursuing extensions to the fast multipole method
(FMM) by Greengard and Rokhlin \cite{Greengard1987}. There exist FMMs
that incorporate periodicity in 3D, e.g.  Kudin \& Scuseria
\cite{Kudin2004} for electrostatics, and FMM-related methods for
Stokes \cite{Greengard2004} in 2D with periodicity. However, the
dominating framework for periodic problems in this setting
incorporates periodicity by using Fourier transforms. The basic
principles for how to proceed in that direction have long been clear,
neatly divided into two stages:
\begin{enumerate}
\item[(a)] Decompose the Stokeslet sum \eqref{eq:stokeslet_sum} into
  rapidly converging parts, e.g. by applying ideas from \emph{Ewald
    summation}.
\item[(b)] Devise a method to reduce the complexity of computing the
  decomposed Stokeslet sums, e.g. by means of FFT-based methods.
\end{enumerate}

The extent to which this has been realized depends on the periodic
structure of the problem. In the fully periodic (3P) setting, various
decompositions exist that clarify $(a)$, and methods with $\O(N\log
N)$ complexity have been developed $(b)$, as we survey in Section
\ref{sec:stokes_ewald_3p}.

In the case of planar periodicity (2P), the picture is much less
clear. Existing decompositions pertaining to $(a)$ are surveyed in
Section \ref{sec:stokes_ewald_2p}, after which we derive an Ewald-type
sum for computing \eqref{eq:stokeslet_sum} (Section
\ref{sec:stokes_ewald_2p_deriv}). A fast, $\O(N\log N)$, method is
presented in Section \ref{sec:fast_method}, which, as far as we know,
is the first reported method that deals with $(b)$ in the 2-periodic
setting.

Interestingly, the fast method we propose $(b)$ is, in some ways,
simpler than the underlying Ewald sum $(a)$ that we derive. As shall
become clear, this has to do with a relationship between the 2- and
3-periodic settings, which we exploit for the fast method.

We demonstrate several appealing characteristics of the proposed
method: spectral accuracy; efficiency in both run-time and memory;
clear error estimation and parameter selection; and a close and
revealing correspondence to methods for the 3P problem. This will, we
hope, facilitate future computational investigations of micro- and
complex flow systems, enabling large systems to be simulated
accurately.

\section{Stokeslet Ewald sum in full periodicity} 
\label{sec:stokes_ewald_3p}

We start by summarizing well-established results for the fully
periodic case. Before interest in solving Stokes equations took off
there was already a large body of work established on a related
problem in electrostatics -- summing Coulomb potentials under
periodicity (solving a Poisson problem, rather than Stokes) -- known
as Ewald summation. The basic principle is that the potential, $\phi
\sim 1/r$, is split into a short range part that is exponentially
decaying, and a long range part that is very smooth (and thus
exponentially convergent in reciprocal space).

Pioneering work by Hasimoto \cite{Hasimoto1959} showed that a
3-periodic vector field $\U^{3P}(\x)$ can be computed in a Ewald-like
manner,
\begin{multline}
  \U^{3P}(\x_m) = \sum_{\p\in \Z^3} \sumn A(\xi, \x_m - \x_n + L\p)\f_n +\\
  +\frac{8\pi}{L^3} \sum_{\k \neq 0} B(\xi,\k) e^{-k^2/4\xi^2}\sumn
  \f_n \expk{(\x_m - \x_n)} - \frac{4\xi}{\sqrt{\pi}}
  \f_m, \label{eq:stokes_ewald_3p}
\end{multline}
where
\begin{align}
  A(\xi, \x) = 2\left( \frac{\xi e^{-\xi^2 r^2}}{\sqrt{\pi} r^2} +
    \frac{\erfc{(\xi r)}}{2 r^3} \right) (r^2 \id + \x\otimes\x) -
  \frac{4\xi}{\sqrt{\pi}} e^{-\xi^2 r^2}\id, \label{eq:hasimoto_real}
\end{align}
with $r := \| \x\|$, and
\begin{align}
  B(\xi, \k) = \left(1 + \frac{k^2}{4\xi^2} \right)
  \frac{1}{k^4}(k^2 \id - \k\otimes\k). \label{eq:hasimoto_fd}
\end{align}
The parameter $\xi>0$, which $\U^{3P}$ is independent of, is known as
the Ewald parameter and controls the rate at which the two sums
converge relative to each other. Note here that the two sums in
\eqref{eq:stokes_ewald_3p} have the desired structure -- the first is
a sum in real space that converges roughly as $e^{-\xi^2 r^2}$, and
the second is a sum in $\k$-space that converges as roughly as
$e^{-k^2/4\xi^2}$.  Other decompositions exist, see Pozrikidis
\cite[Sec. 3.1, 4.1]{Pozrikidis1996}.

\subsection{Fast methods in full periodicity} \label{sec:3p_fast_meth}

The $\sim 1/r$ convergence of the original Stokeslet sum
\eqref{eq:stokeslet_sum} is \emph{vastly} improved by the Ewald method
\eqref{eq:stokes_ewald_3p}. However, the complexity of computing the
Ewald sum \eqref{eq:stokes_ewald_3p} is still severely limiting. There
are two reasons for this: First, summing \eqref{eq:stokes_ewald_3p}
for all $\x_m$ has $\O(N^2)$ complexity. Secondly, the constant hidden
in the formal complexity can be \emph{very} large; in fact, it grows
cubically with higher accuracy.

Faster methods for the corresponding Poisson problem in electrostatics
and molecular simulation have been around for three decades, following
work by Hockney \& Eastwood \cite{Hockney1981}; see the survey by
Deserno \& Holm \cite{Deserno1998}, or recent work by the present
authors \cite{Lindbo2011}.

Such methods have been adapted for the 3P Stokes Ewald sum
\eqref{eq:stokes_ewald_3p}, staring with Sierou \& Brady
\cite{Sierou2001} (embedded in the framework of ``Stokesian
dynamics''). Their method is based on the \emph{Particle Mesh Ewald}
(PME) method by Darden et al. \cite{Darden1993}. Saintillan
et al. \cite{Saintillan2005}, in their method for sedimenting fibers
in Stokes flow, base a fast method for the Stokeslet sum on a
refinement of the PME method, known as \emph{Smooth Particle Mesh
  Ewald} (SPME) \cite{Essmann1995}. The SPME method is known to be
more accurate than its predecessor, though still of polynomial
order. Recognizing that the exponentially fast convergence of the
underlying Ewald sums is lost when a traditional PME approach is used,
the present authors presented a spectrally accurate PME-type method
for Stokes in \cite{Lindbo2010}.

\section{2-periodic Stokeslet Ewald sum} 
\label{sec:stokes_ewald_2p}

Perhaps counter to intuition, one should not expect a method for
2-periodic systems to follow by elementary manipulations of the 3P
decomposition \eqref{eq:stokes_ewald_3p} -- to the contrary, one is
best advised to start anew from the Stokeslet sum
\eqref{eq:stokeslet_sum}. It is also instructive to review the
corresponding transition from 3P to 2P in the electrostatics setting,
as we do in \cite{Lindbo2011a}. One finds that consolidation, on the
level of agreeing on a preferred decomposition, has yet to happen
(cf. e.g. the 2P Ewald sum in Gryzbowski et al. \cite{Grzybowski2000}
and a survey of non-Ewald methods by Mazars \cite{Mazars2005}), and
that fast PME-type methods are considerably less mature than their 3P
counterparts, though we hope that \cite{Lindbo2011a} contributes in
this direction.

There does exist 2P Ewald decompositions for Stokes -- notably in
Pozrikidis \cite{Pozrikidis1996}, relating to earlier work by Ishii
\cite{Ishii1979}. While it should be emphasized that
\cite{Pozrikidis1996} is among the most valuable references in the
present context, we find the results given therein lacking in two
respects: First, it turns out to be quite straight-forward to derive
and present a 2P Stokes Ewald sum in explicit form, whereas Pozrikidis
\cite[Sec. 2.2]{Pozrikidis1996} is content with giving a ``generating
function'' and a differential operator. Secondly, and more
importantly, the results on offer in \cite{Pozrikidis1996} do not, at
least to us, suggest how a fast PME-type method could be developed.

Such a fast method being our objective, we now set out to derive a 2P
Ewald sum for Stokes in a suitable way. To give an overview, we start
by deriving a pure $\k$-space solution to a 2-periodic Stokes problem
that decays as $z\rightarrow\pm\infty$, which we then decompose using
the ``screening function'' \eqref{eq:screening_fcn} proposed by
Hernandez-Ortiz et al. \cite{Hernandez-Ortiz2007} to generate the
Hasimoto decomposition
\eqref{eq:stokes_ewald_3p}-\eqref{eq:hasimoto_fd} in the 3-periodic
setting. We then consider the family of solutions obtained by adding a
piecewise linear function, showing that this admits an end-result
which is differentiable. Under certain conditions, we show that these
smooth solutions have finite limits as $z\rightarrow\pm\infty$, as
expected physically. The resolution of the smoothness/singularity
issue is quite compactly discussed in related work
\cite[pp. 83-84]{Pozrikidis1996}, \cite[p. 676]{Ishii1979}; this may
suit some readers more than others.

\subsection{2P Stokeslet Ewald derivation: preliminaries}

\begin{figure}
  \centering
  \includegraphics{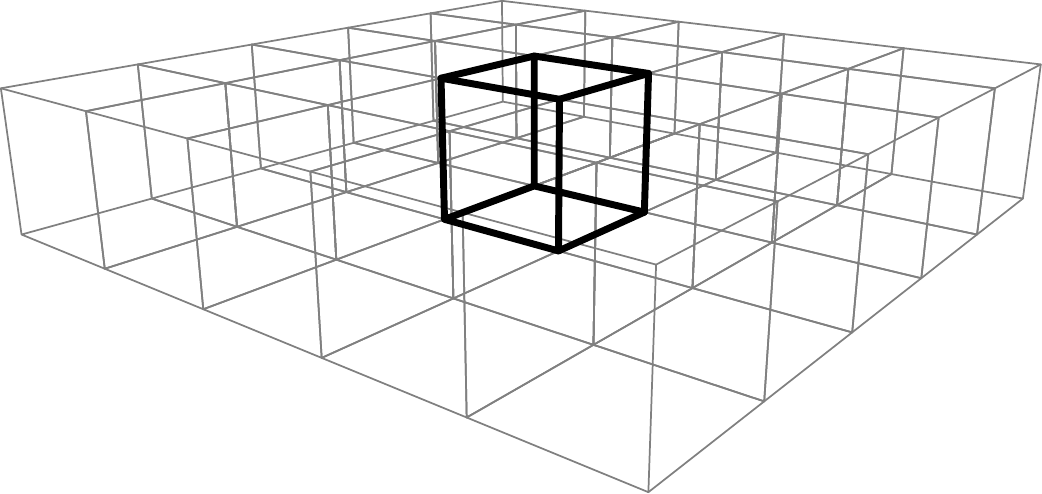}
  \caption{Planar periodicity (2P): Primary cell replicated infinitely
    in the plane. }
  \label{fig:2p}
\end{figure}

\subsubsection{A mixed Fourier transform} \label{sec:prelim}

The conventions for Fourier transform pairs used in the present work
are given in Appendix \ref{app:ft_defn}. As in \cite{Lindbo2011a}, we
start with observing that the periodic structure of a 2-periodic
function $h(\x)$ implies that its spectral representation will be
\emph{mixed} in the following sense: Let $h(\x) = h(\r,z)$ be periodic
in the $(x,y)$-plane and ``free'' in $z$, i.e. $\r:=(x,y) \in \Omega =
[0,L)^2$ and $z\in \R$. Then a Fourier representation of $h$ is
\begin{align}
  h(\r,z) = \frac{1}{2\pi}\sum_\k \int_{-\infty}^\infty
  \hat{h}(\k,\kappa) e^{i\k\cdot \r} e^{i\kappa z} \d
  \kappa, \label{eq:mixed_ft}
\end{align}
where $\k\in 2\pi\Z^2/L$ and $\kappa\in\R$. We shall assume that
$h(\r, z \rightarrow \pm \infty)$ decays faster than any inverse
power of $z$. In this setting, \eqref{eq:mixed_ft} exists. Also under
the present assumptions, 2P versions of several familiar results from
spectral analysis hold, such as the Poisson summation formula,
Parseval/Plancherel's theorems and the convolution theorem, see
\cite{Lindbo2011a}.

\subsubsection{Solution of a 2-periodic Stokes  problem} 
\label{sec:unscreened_soln}

Before we address Ewald summation for the 2P Stokes problem, we need
to establish (slowly converging) solutions of the original problem
that, crucially, vanishes as $z\rightarrow \pm \infty$. Consider the
Stokes equations,
\begin{align*}
  -\nabla p(\x) + \mu \Delta \U(\x) + \f(\x) = 0 \\
  \nabla \cdot \U(\x) = 0
\end{align*}
under 2-periodic boundary conditions with respect to
$\Omega$. Presently, assume that the source term, $\f$, has a mixed
Fourier transform, $\hat{\f}(\k,\kappa)$, and expand $\U$ and
$\q:=\nabla p$ likewise,
\begin{align}
  \U(\x) &= \U(\r,z) = \frac{1}{2\pi}\sum_\k \int_\R \hU(\k,\kappa)
  e^{i\k\cdot \r} e^{i\kappa z} \d \kappa \label{eq:u_mixed_ft}\\
  \nabla p(\x) &= \nabla p(\r,z) = \frac{1}{2\pi}\sum_\k \int_\R
  \hat{\q}(\k,\kappa) e^{i\k\cdot \r} e^{i\kappa z} \d
  \kappa. \nonumber
\end{align}
Inserting into Stokes equations, with $\kf:=(\k,\kappa)\in
\{2\pi\Z^2/L\}\times\R$,
\begin{align*}
  \hat{\q} + \mu \nkf^2 \hU &= \hat{\f}\\
  i \kf\cdot \hat{\U}&=0.
\end{align*}
As $\U$ should vanish at infinity we may omit summing over $\k=0$,
because $\hat{\U}(0,\kappa)=0$ is implied. Eliminating the pressure
gradient above, $\hat{\q}(\k,\kappa) = (\kf\otimes\kf
)\hat{\f}(\k,\kappa)/\nkf^2$ (since $\q$ is a gradient $\hat{\q}(\k)$
is parallel to $\k$), gives
\begin{align}
  \hat{\U}(\kf)= \hU(\k,\kappa) = \frac{1}{\mu}\left(
    \frac{\id}{\nkf^2} - \frac{\kf \otimes \kf}{\nkf^4}\right)
  \hat{\f}(\k,\kappa), \quad \k\neq 0. \label{eq:u_hat}
\end{align}
Hence, in light of \eqref{eq:u_mixed_ft},
\begin{align}
  \U(\r,z) = \frac{1}{2\pi\mu} \sum_{\k\neq 0} \int_\R \left(
    \frac{\id}{\nkf^2} - \frac{\kf \otimes \kf}{\nkf^4}\right)
  \hat{\f}(\k,\kappa) e^{i\k\cdot \r} e^{i\kappa z}
  \d\kappa. \label{eq:u_fd_full}
\end{align}
Recalling $\f(\x)$ from \eqref{eq:f_deltas}, we absorb a factor
$8\pi\mu$ into the coefficients $\f_n$ for convenience and conformity
with convention. With that, \eqref{eq:u_fd_full} becomes
\begin{align*}
  \tU(\r,z)= \frac{4}{L^2} \sum_{\k\neq 0} \int_\R \left(
    \frac{\id}{\nkf^2} - \frac{\kf \otimes \kf}{\nkf^4}\right)
  \sumn \f_n e^{i\k\cdot (\r-\r_n)} e^{i\kappa (z-z_n)}
  \d\kappa,
\end{align*}
where the integrals are computable. We obtain (see Appendix
\ref{app:non-screened_u_integrals}):
\begin{align}
  \tU(\r,z) = \frac{4}{L^2} \sum_{\k\neq 0} \sumn \tQ(\k,z-z_n) \f_n
  e^{i\k\cdot(\r-\r_n)}, \label{eq:unscreened_u}
\end{align}
where
\begin{align}
  \tQ(\k,z)
  =\frac{e^{-\nk |z| }}{\nk} &
  \begin{bmatrix}
    \pi -\frac{(\pi \nk |z|+\pi ) k_1^2}{2 \nk ^2} & -\frac{\pi (\nk
      |z|+1) k_1 k_2}{2 \nk ^2} &
    -\frac{1}{2} i \pi  z k_1 \\
    -\frac{\pi (\nk |z|+1) k_1 k_2}{2 \nk ^2} & \pi -\frac{(\pi \nk
      |z|+\pi ) k_2^2}{2 \nk ^2} &
    -\frac{1}{2} i \pi  z k_2 \\
    -\frac{1}{2} i \pi z k_1 & -\frac{1}{2} i \pi z k_2 & \frac{1}{2}
    (\pi \nk |z|+\pi )
  \end{bmatrix}. \label{eq:unscreened_Q}
\end{align}
Note that $\tU$ is well-defined, but, as will be discussed in Section
\ref{sec:2p_direc_fd}, fails to be differentiable in an $(x,y)$-plane
around each $z_n$.

\subsection{2P Stokeslet Ewald derivation} 
\label{sec:stokes_ewald_2p_deriv}

A fruitful and flexible approach to deriving Ewald sums is to compute
convolutions of source term with a so called screening function --
this is the classical approach for the 3P Laplace (electrostatics)
problem. In planar periodicity the singularities encountered are more
severe, and the appropriate behavior in the limit
$z\rightarrow\pm\infty$ not as straight forward as the corresponding
``tin foil'' (or, for Stokes,``pressure gradient counters net force'')
conditions that enter 3P derivations. Our approach here is to
construct a decomposition of \eqref{eq:unscreened_u} and then address
the regularity issue.

By ``screening function'' we mean any normalized function,
$\|\gamma(\x)\|_2 = 1$, that decays smoothly away from zero, such that
a decomposition,
\begin{align*}
  \f(\x) &= ((\delta-\gamma)* \f)(\x) + (\gamma*\f)(\x) = \f(\x)
  -\f_\gamma(\x) + \f_\gamma(\x),
\end{align*}
where $\delta(\x)$ denotes the Dirac measure on $\R^3$ and
$\f_\gamma(\x) := (\f*\gamma)(\x)$, is well defined. With this,
clearly,
\begin{align}
  \U(\x) = \frac{1}{8\pi\mu}\int_{\R^3} S(\x-\y) \left( \f(\y) -
    \f_\gamma(\y) \right) \d \y + \frac{1}{8\pi\mu}\int_{\R^3} S(\x-\y) 
  \f_\gamma(\y) \d\y. \label{eq:u_decomp}
\end{align}
As noted, the standard derivation for the 3P Laplace case is to take
$\gamma$ as a pure Gaussian, compute the first convolution directly
and treat the second term in reciprocal space. Interestingly, in a
short paper sparse on details \cite{Hernandez-Ortiz2007},
Hernandez-Ortiz et al. give a screening function,
\begin{align}
  \gamma(r) = \frac{\xi^3}{\sqrt{\pi^3}} e^{-\xi^2 r^2}\left(
    \frac{5}{2} - \xi^2 r^2 \right), \label{eq:screening_fcn}
\end{align}
that generates exactly the 3P Hasimoto decomposition
\eqref{eq:stokes_ewald_3p}-\eqref{eq:hasimoto_fd}. 

\subsubsection{The real-space sum}

The first term in \eqref{eq:u_decomp} can be computed directly, as is
standard in this context (though the calculations are laborious), and
one finds that
\begin{align}
  \int_{\R^3} S(\x-\y) \left( \f(\y) - \f_\gamma(\y)
  \right) \d \y = \sumn \sump{2} A(\xi, \x - \x_n +
  L\p)\f_n, \label{eq:ewald_sum_rs_non_lim}
\end{align}
where $A$ is the same (though summed over a two-dimensional lattice
this time) \emph{vis-a-vis} the 3P decomposition
\eqref{eq:hasimoto_real}. 

Note that a particle does not itself contribute to the potential
field, or flow, it experiences, so it is natural to simply drop the
term in \eqref{eq:ewald_sum_rs_non_lim} that corresponds to $\p=0$
when $m=n$.  However, by the decomposition \eqref{eq:u_decomp}, part
of the contribution that we're trying to remove has gone into the
second term. By computing the difference between the free space
Stokeslet and the real-space term, we can find this contribution. This
is to be subtracted off, so we do it with the opposite sign:
\begin{align*}
  \lim_{\|\x\|\rightarrow 0} \left( A(\x,\xi) - S(\x) \right) =
  \lim_{\| \x\|\rightarrow 0} \left(
    -(\alpha+\erf(\xi\|\x\|))\frac{\id}{\|\x\|} +
    (\alpha-\erf(\xi\|\x\|))\frac{\x\otimes\x}{\|\x\|^3} \right) = -
  \frac{4\xi}{\sqrt{\pi}} \id,
\end{align*}
where $\alpha:=2\xi\pi^{-1/2}\|\x\|e^{-\xi^2\|\x\|^2}$.  By
convention, this term is denoted ``self interaction'' which can lead
to some confusion, since it's the term removing self interaction that
is included in the $\k$-space sum. Nonetheless, we have arrived at two
parts of the 2P Stokeslet Ewald sum,
\begin{align}
  \U^R(\x_m) &:= \sumn \sump{2}^* A(\xi, \x_m - \x_n + L\p)\f_n
  \label{eq:ewald_sum_rs}\\
  \Uself(\x_m) &:= -\frac{4\xi}{\sqrt{\pi}} \f_m. \label{eq:ewald_sum_self}
\end{align}
The notational inconvenience $*$, signifying the omission of the term
$\{\p=0,n=m\}$ in \eqref{eq:ewald_sum_rs}, is standard.

\subsubsection{The $\k$-space sum} \label{sec:2p_direc_fd}

Turning to the second term in the Stokeslet decomposition
\eqref{eq:u_decomp}, first note that both terms in \eqref{eq:u_decomp}
are solutions to Stokes equations. Moreover, both $\f_\gamma$ and
$\f-\f_\gamma$ are admissible under the assumptions of Section
\ref{sec:unscreened_soln}. The Fourier transform of $\gamma$ over
$\R^3$ is
\begin{align*}
  \hat{\gamma}(k) = \left(1+\frac{k^2}{4\xi^2}\right)e^{-k^2/4\xi^2}.
\end{align*}
In light of \eqref{eq:u_fd_full} we let $\f_\gamma=\f*\gamma$ generate
a function $\U^F$ that satisfies $\U^F\rightarrow 0$ as
$z\rightarrow\pm\infty$,
\begin{align}
  \U^F(\r,z) &= \frac{1}{2\pi\mu}\sum_{\k\neq 0} \int_\R \left(
    \frac{\id}{\nkf^2} - \frac{\kf \otimes \kf}{\nkf^4}\right)
  \widehat{\f_\gamma}(\k,\kappa) e^{i\k\cdot \r} e^{i\kappa z}
  \d\kappa\nonumber \\
  & = \frac{4}{L^2}\sum_{\k\neq 0} \int_\R \left(
    \frac{\id}{\nkf^2} - \frac{\kf \otimes \kf}{\nkf^4}\right)
  \left(1+\frac{\nkf^2}{4\xi^2}\right)e^{-\nkf^2/4\xi^2} \sumn f_n
  e^{i\k\cdot (\r-\r_n)}
  e^{i\kappa (z-z_n)} \d\kappa \label{eq:ewald_sum_fd_int} \\
  &= \frac{4}{L^2}\sum_{\k\neq 0} \int_\R B(\kf) e^{-\nkf^2/4\xi^2}
  \sumn \f_n e^{i\k\cdot (\r-\r_n)} e^{i\kappa(z-z_n)}
  \d\kappa,\nonumber
\end{align}
where we have used Poisson summation and grouped similar Fourier
coefficients.

For future reference, we note that this expression is the starting
point from which we develop a fast PME-type method. Given the present
mixed periodic setting, the close correspondence between
\eqref{eq:ewald_sum_fd_int} and \eqref{eq:hasimoto_fd} is entirely
expected, and illuminating \emph{per se}.

As it transpires in Appendix \ref{app:screened_u_integrals}, the
integrals above,
\begin{align}
  Q(\k,z) := e^{-\nk^2/4\xi^2} \int_\R \left(
    \frac{\id}{\nkf^2} - \frac{\kf \otimes \kf}{\nkf^4}\right)
  \left(1+\frac{\nkf^2}{4\xi^2}\right) e^{-\kappa^2/4\xi^2}
  e^{i\kappa z}\d \kappa, \label{eq:Q}
\end{align}
can be computed. For clarity of notation is is useful to let
\begin{align*}
  Q(\k,z) = \QI(\k,z) + \Qkk(\k,z),
\end{align*}
where the meaning of the superscripts is implied from the terms in the
first factor under the integral in \eqref{eq:Q}. With this, and the
computations in Appendix \ref{app:screened_u_integrals}, it follows
that we may write $\U^F$ as a sum,
\begin{multline}
  \U^F(\r_m, z_m) = \frac{4}{L^2} \sum_{\k\neq 0} \sumn \bigg(
  \big(\QI(\k,z_{mn}) + \Re \Qkk(\k,z_{mn}) \big)
  \cos(\k\cdot\r_{mn} ) -\\ + \Im \Qkk(\k,z_{mn}) \sin(\k\cdot
  \r_{mn} ) \bigg)\f_n,
  \label{eq:ewald_sum_fd}
\end{multline}
where
\begin{align}
  \QI(\k,z) &= 2 \left( \frac{J^0_{0}(z)}{4\xi^2}  + 
    J^1_{0}(\nk,z)\right) \id
\end{align}
and
\begin{align} 
  \Qkk(\k,z) &= -2\left[
    \begin{array}{ccc}
      k_1^2 \left(\frac{J^0_{1}}{4 \xi ^2}+J^0_{2}\right) & 
      k_1 k_2 \left(\frac{J^0_{1}}{4 \xi ^2}+J^0_{2}\right) & 
      k_1 \left(\frac{i K^1_{1}}{4 \xi ^2}+i K^1_{2}\right) \\
      k_1 k_2 \left(\frac{J^0_{1}}{4 \xi ^2}+J^0_{2}\right) & 
      k_2^2 \left(\frac{J^0_{1}}{4 \xi ^2}+J^0_{2}\right) & 
      k_2 \left(\frac{i K^1_{1}}{4 \xi ^2}+i K^1_{2}\right) \\
      k_1 \left(\frac{i K^1_{1}}{4 \xi ^2}+i K^1_{2}\right) & 
      k_2 \left(\frac{i K^1_{1}}{4 \xi ^2}+i K^1_{2}\right) & 
      \frac{J^2_{1}}{4 \xi ^2}+J^2_{2}
    \end{array}
  \right](\nk,z).
\end{align}
The terms $J^p_q$ and $K^p_q$ are short-hand for the various scalar
integrals that can be identified by studying the integrand in $Q$. With
\begin{align}
  \lambda  &:= e^{-k^2/4\xi^2 - \xi^2 z^2}\\
  \theta_+ &:= e^{kz} \erfc\left( \frac{k}{2\xi} + \xi z\right)\\
  \theta_- &:= e^{-kz} \erfc\left( \frac{k}{2\xi} - \xi z\right)
\end{align}
we can write down the computed integrals as
\begin{align}
  J^0_0(z,k)  &= \sqrt{\pi} \lambda \xi \label{eq:J00}\\
  J^0_1(z,k)  &= \frac{\pi  \left(\theta _-+\theta _+\right)}{4 k}\\
  J^0_2(z,k) &=\frac{\sqrt{\pi } \lambda }{4 k^2 \xi }+\pi
  \left(\frac{\theta _-+\theta _+}{8 k^3}+\frac{\left(\theta _--\theta
        _+\right) z}{8 k^2}-\frac{\theta _-+\theta _+}{16 k \xi ^2}\right)\\
  J^2_1(z,k) &= \frac{1}{4} \pi \left(-\theta _--\theta _+\right) k+
  \sqrt{\pi } \lambda  \xi\\
  J^2_2(z,k) &=\pi \left(\frac{\left(\theta _-+\theta _+\right) k}{16
      \xi ^2}+\frac{\theta _-+\theta _+}{8 k}+\frac{1}{8} \left(\theta
      _+-\theta _-\right)
    z\right)-\frac{\sqrt{\pi } \lambda }{4 \xi }\\
  K^1_1(z,k) &= \frac{\pi(\theta_- - \theta_+ )}{4}\\
  K^1_2(z,k) &= \pi\left( \frac{\theta_+ - \theta_-}{16\xi^2} +
    \frac{(\theta_- + \theta_+)z}{8k}\right)\label{eq:K12}.
\end{align}
Despite some difficulty of notation, the sum \eqref{eq:ewald_sum_fd}
is straight-forward to evaluate.

Now, recall that we are presently aiming for a decomposition of the
pure $\k$-space solution \eqref{eq:unscreened_u}, wherein the $\k=0$
term was dropped to ensure decay as $z\rightarrow\pm\infty$. However,
by the decomposition \eqref{eq:u_decomp}, part of the $\k=0$ mode has
gone into the real-space sum \eqref{eq:ewald_sum_rs} and must be
subtracted off. Computing \eqref{eq:unscreened_u} explicitly has
provided a $\k$-space view of the real-space sum, so that the term to
be removed can be extracted as
\begin{align*}
  \U^*(z) =  \lim_{\k\rightarrow 0} \left( \hU(\k) - \hU^F(\k) \right) = 
  \lim_{\k\rightarrow 0} \frac{4}{L^2}\left( \sumn (\tQ(\k,z-z_n) -
    Q(\k,z-z_n))\f_n\right).
\end{align*}
Computing the desired limit, one finds
\begin{align*}
  \lim_{\k\rightarrow 0} \left( Q(\k,z) - \tQ(\k,z)
  \right) =\left(
    \begin{array}{ccc}
      a(z) & 0 & 0 \\
      0 & a(z) & 0 \\
      0 & 0 & 0
    \end{array}
  \right),
\end{align*}
where
\begin{align}
  a(z) := \pi |z|-\pi  z \erf(z \xi )-\frac{e^{-z^2 \xi ^2} 
    \sqrt{\pi }}{2 \xi }, \label{eq:Q_lim}
\end{align}
so that
\begin{align*}
  \U^*(z) = \frac{4}{L^2}\sumn a(z-z_n)\id_2 \f_n, \quad
  \id_2 := \begin{bmatrix}
    1 & 0 & 0\\
    0 & 1 & 0\\
    0 & 0 & 0
  \end{bmatrix}.
\end{align*}
We now have the desired decomposition of \eqref{eq:unscreened_u},
\begin{align*}
  \tU(\r_m,z_m) = \U^R(\r_m, z_m) + \U^F(\r_m,z_m) - \U^*(z_m) +
  \Uself.
\end{align*}
Naturally, the non-smoothness in \eqref{eq:unscreened_u} is present
here (in the term $\U^*$). However, we can recover a smooth solution
by relaxing the restriction that $\U\rightarrow 0$ as
$z\rightarrow\pm\infty$. To any $\U$ that satisfies the Stokes
equation \eqref{eq:stokes_eq} we can add a linear function. By that
token, and in close analogy to how the 2P Ewald sum for electrostatics
was obtained in \cite{Lindbo2011a}, we subtract a piecewise linear
function,
\begin{align}
  \U^{F,\k=0}(\x_m) &= -\frac{4\pi}{L^2} \sumn |z_m-z_n| \id_2 \f_n - \U^*
  \nonumber\\
  & = -\frac{4}{L^2} \sumn \left( \pi (z_m - z_n) \erf\big( (z_m-z_n)
    \xi \big) + \frac{\sqrt{\pi}}{2\xi} e^{-(z_m-z_n)^2 \xi^2}\right)
  \id_2\f_n. \label{eq:ewald_sum_k0}
\end{align}
With this, the derivation of the 2P Stokes Ewald sum is complete. We
have that
\begin{align}
  \U(\x_m) = \U(\r_m, z_m ) = \U^R(\r_m, z_m) + \U^F(\r_m,z_m) +
  \Ufz(z_m) + \Uself, \label{eq:stokeslet_ewald_sum_2p}
\end{align}
from \eqref{eq:ewald_sum_rs}, \eqref{eq:ewald_sum_fd},
\eqref{eq:ewald_sum_k0} and \eqref{eq:ewald_sum_self}. These are nice
and exponentially convergent sums, but, as discussed in the
introduction, the complexity of evaluating them for a large system is
debilitating. Note that $\Ufz$ contains terms that look like an
unbounded shear flow in $z$. However, if the coefficients $\f_n$ sum
to zero in the first two components, $\Ufz$ tends to finite limits as
$z\rightarrow\pm\infty$. As in the electrostatics case, one can show
that
\begin{align*}
  \left( \lim_{z\rightarrow\infty} \U(\r,z) -
    \lim_{z\rightarrow-\infty} \U(\r,z) \right) = \frac{8\pi}{L^2}\sumn z_n 
  \f_n \id_2, \quad \text{if} \quad \sumn (\f_n)_j = 0, \,j=1,2.
\end{align*}
That is, in the planar directions $(x,y)$ the flow sees a transition
(across $z\in(0, L)$ continuously) proportional to the dipole moment
in z.

\subsection{Truncation error estimates for 2P Stokeslet Ewald sums}
\label{sec:trunc_est}

It's noteworthy that the usefulness of Ewald sums (cf. Section
\ref{sec:param_selection}, on parameter selection) rests on the
availability of truncation error estimates. Therefore, we need to
endow the 2P Stokeslet Ewald sums, \eqref{eq:ewald_sum_rs} and
\eqref{eq:ewald_sum_fd}, with such estimates.

To this end, let the real-space sum \eqref{eq:ewald_sum_rs} be
truncated such that only interactions between particles (including
periodic images) within a distance $r_c$ from each other are
included. That is, let
\begin{align*}
  \U_{r_c}^R(\x_m) &:= \sumn \sum_{\p\in \Z^2}^* \mathbf{1}_{r_c}(\|\x_m
  - \x_n + \tau(\p)\|)A(\xi, \x_m - \x_n + \tau(\p))\f_n,
\end{align*}
where $\mathbf{1}_{r_c}(r)$ denotes the indicator function which is
one if $r\leq r_c$ and zero otherwise. Estimating $\| \U^R -
\U_{r_c}^R\|$ is most tractable in the RMS norm,
\begin{align}
  e_{\rms}:= \sqrt{\frac{1}{N} \sumn (\U(\x_n) - \U_{*}(\x_n)
    )^2}. \label{eq:err_rms}
\end{align}
In the electrostatic case, Kolafa \& Perram \cite{Kolafa1992} have
derived famous truncation error estimates for randomly scattered
particles in RMS norm. However, the diagonal terms in $A$ are $A_{jj}
=\erfc(\xi r)/r - 2\xi \exp(-\xi^2 r^2)/\sqrt{\pi}$, whereas in the
Laplace case only the former (and smaller) term is present. On the
other hand, if we disregard the off-diagonal terms in $A$, the key
step in \cite[Appendix A, Eq. (14)]{Kolafa1992} becomes tractable for
the Stokeslet, and we get estimates for components $j=1,2,3$,
\begin{align}
  (e^R_{\rms,j})^2 &\approx \frac{Q_j}{L^3} \int_0^{2\pi} \int_0^\pi
  \int_{r_c}^\infty A_{jj}^2 r^2 \sin(\theta) \d r
  \d\theta \d\phi \nonumber\\
  & = \frac{Q_j}{L^3}\left(4 r_c\left( e^{-2\xi^2 r_c^2} - \pi
      \erfc(\xi r_c)^2 \right) + \frac{\sqrt{2\pi}}{\xi}\erfc(\sqrt{2}
    \xi r_c)\right), \label{eq:rs_rms_trunc_est}
\end{align}
where $Q_j:=\sumn (f_j)_n^2$. In what follows, we suppress the vector
index $j$.

\begin{figure}
  \centering
  \includegraphics{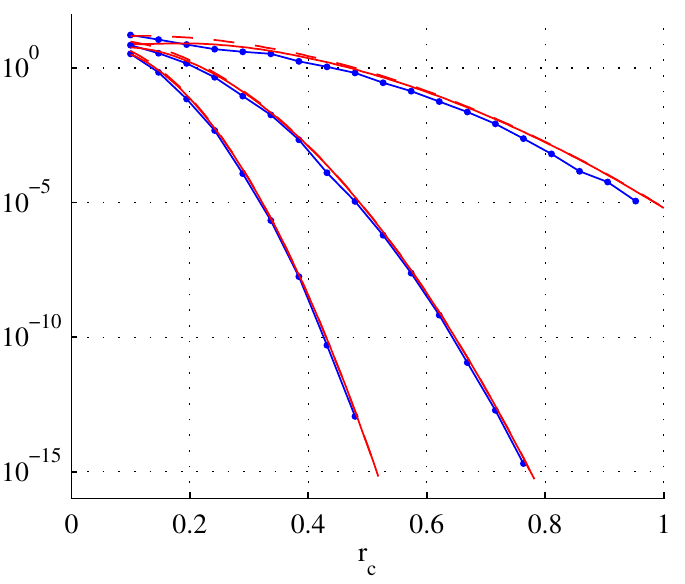}
  \caption{Real-space sum truncation error \eqref{eq:err_rms} and
    estimate \eqref{eq:rs_rms_trunc_est} of $x$-component, in
    RMS-norm, as a function of truncation radius. $N=10000$,
    $\Omega=[0, 1)^3$, $\xi=4,8,12$ (top to bottom). As dashed,
    simplified estimate \eqref{eq:err_R_rms_simpl}, hardly
    distinguishable. }
  \label{fig:rsrc_trunc}
\end{figure}

The \emph{estimate} \eqref{eq:rs_rms_trunc_est} is to be treated as
such -- it does not include the full operator and it is statistical,
the latter meaning that it's only valid if $\x_m$ is randomly
distributed. None the less, \eqref{eq:rs_rms_trunc_est} is very
predictive within its domain, as we illustrate in Figure
\ref{fig:rsrc_trunc}. Here we have $N=10000$, $\x_m$ randomly from a
uniform distribution over $\Omega=[0, 1)^3$, $\f_m$ from a uniform
distribution centered at zero and the RMS average was formed over 30
random $\x_m$. Finally, for selecting parameters one would ideally
like to solve $e^R_{\rms}(\xi)=\epsi$ for $\xi$, but this does not
appear tractable. Rather, we series expand the error estimate for
large $\xi r_c$, obtaining that
\begin{align}
  e^R_{\rms} \approx 2 \sqrt{\frac{Q r_c}{L^3} }e^{-\xi^2
    r_c^2}, \label{eq:err_R_rms_simpl}
\end{align}
and consequently, assuming that $r_c \geq L^3\epsi^2/(4 Q)$,
\begin{align*}
  \xi(r_c,\epsi) \approx \frac{1}{r_c} \sqrt{\log
    \left(\frac{2}{\epsi}\sqrt{\frac{Q r_c}{L^3}}\right)}.
\end{align*}
The agreement between computed errors and the simplified estimate
\eqref{eq:err_R_rms_simpl} is illustrated in Figure
\ref{fig:rsrc_trunc}.

\begin{figure}
  \centering
  \includegraphics{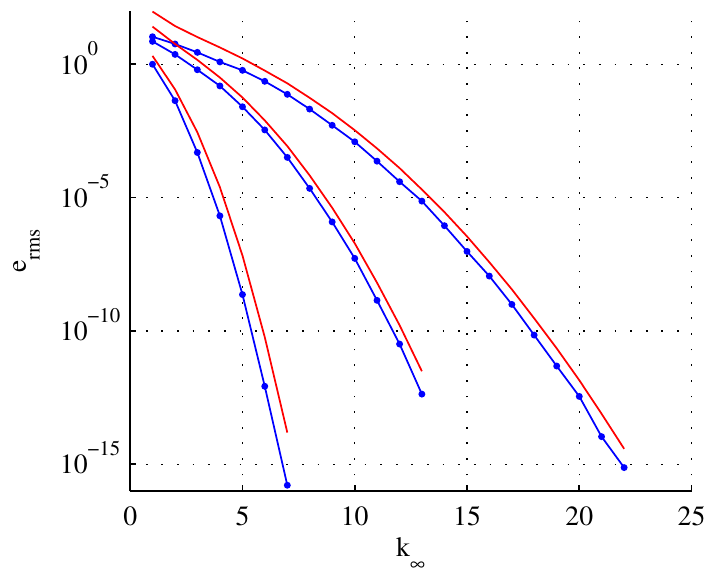}
  \includegraphics{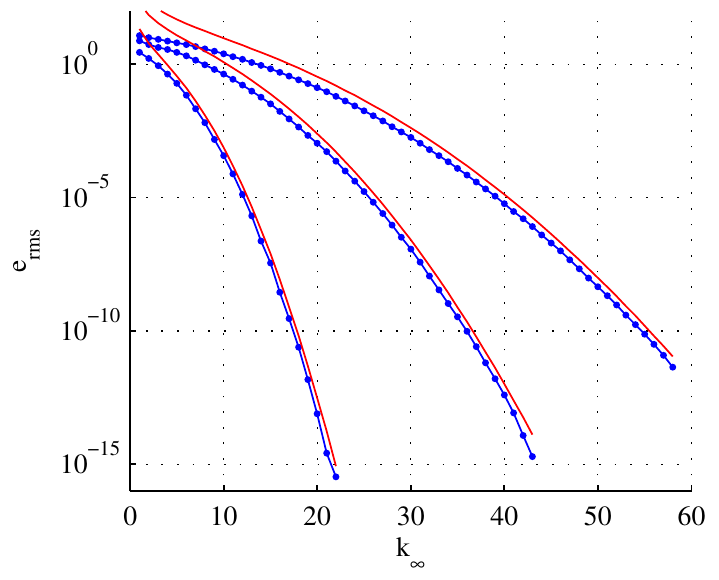}
  \caption{Truncation error of $\k$-space 2P Stokeslet Ewald sum
    \eqref{eq:ewald_sum_fd} in RMS norm \eqref{eq:err_rms} vs. number
    of modes $\kmax$, i.e. $\|\k\|\leq 2\pi \kmax/L$. Left:
    $L=1,N=400$. Right: $L=3, N=200$. In both panels, $\xi=4,8,12$
    (left to right), and (solid line) heuristic error estimate
    \eqref{eq:err_F_rms_heuristic}.}
  \label{fig:fd_trunc}
\end{figure}

For the $\k$-space sum \eqref{eq:ewald_sum_fd}, truncated as
$\|\k\|\leq 2\pi\kmax/L$, it's harder to follow Kolafa \& Perram
\cite{Kolafa1992} and derive a corresponding estimate. Rather than to
ignore the issue altogether, which would break the way parameters are
selected (cf. Section \ref{sec:param_selection}), we take a heuristic
approach. Based on existing results in the RMS setting, it's natural
to suppose that an estimate of the form $e^F_{\rms} \approx C \sqrt{Q}
\kmax^a \xi^b e^{-(\frac{\pi\kmax}{\xi L})^2}$, for particular $a,b$,
can be of value. A few trial runs quite clearly suggest that good
agreement is obtained if $a=-3/2$ and $b=3$. Moreover, we let $C = L^2
\pi^{-4}$, obtaining a heuristic, or practical, error estimate for the
truncation error of the $\k$-space sum \eqref{eq:ewald_sum_fd},
\begin{align}
  e^F_{\rms} \approx \sqrt{Q} \frac{\xi^3 L^2}{\pi^4 \kmax^{3/2}}
  e^{-(\frac{\pi \kmax}{\xi L})^2}. \label{eq:err_F_rms_heuristic}
\end{align}
In Figure \ref{fig:fd_trunc} we give results that demonstrate that
this estimate captures the truncation error well for a range of
parameters. Here, $\xi=4,8,12$ and $L=1, 3$, $N=400,200$ and the
RMS-average is formed over a random set of 30 points\footnote{The
  small systems used to evaluate the estimate
  \eqref{eq:err_F_rms_heuristic} are small -- limited by the
  computational cost of evaluating \eqref{eq:ewald_sum_fd} for large
  $\xi$.}. We observe satisfying agreement, and, as appropriate, that
the estimate is somewhat conservative.

Pertaining to both the real- and $\k$-space sums, the RMS measure will
underestimate errors if particles are not randomly scattered. In that
case, it becomes appropriate to consider $\infty$-norm estimates. The
essential property, that the sums converge at least as fast as
$\exp(-r_c^2 \xi^2)$ and $\exp(-(\pi\kmax/(\xi L))^2)$ respectively,
still holds, though obtaining reliable estimates with detailed
constants, as in the RMS case, is bound to pose a challenge.

\section{Fast methods for 2P Stokes Ewald sums} 
\label{sec:fast_method}

In Section \ref{sec:3p_fast_meth} we briefly surveyed fast, $\O(N\log
N)$, methods for the fully periodic (3P) problem, and we shall adhere
to same framework under 2P:
\begin{itemize}
\item[a)] The real-space sum \eqref{eq:ewald_sum_rs} can be made
  arbitrarily cheap to compute by choosing the Ewald parameter, $\xi$,
  sufficiently large (cf. Section \ref{sec:fast_rs_meth});
\item[b)] by using fast Fourier transform (FFT), the (non-singular)
  $\k$-space sum \eqref{eq:ewald_sum_fd} can be evaluated in linear
  time (independently of $\xi$);
\item[c)] the (2P-only) singularity contribution
  \eqref{eq:ewald_sum_k0} is easily dealt with as a one-dimensional
  interpolation problem.
\end{itemize}
It is really item $(b)$ that is the heart of the matter, that and the
glue that hold the parts together: parameter estimation. In the 3P
setting one starts from the Ewald sum (the second term in
\eqref{eq:stokes_ewald_3p}, in the Stokes case), and the steps to
obtain a fast, FFT-based, method are quite intuitive. That is not to
say that the matter is simple; a rather large debt of insight is owed
to the early pioneers in the field, such as Hockney \& Eastwood
\cite{Hockney1981} and Darden et al. \cite{Darden1993}.

Attempts to do the same thing in planar periodicity immediately run
aground on the algebraic structure of the corresponding $\k$-space
Ewald sum \eqref{eq:ewald_sum_fd} (cf. e.g. Grzybowski
et al. \cite{Grzybowski2000} for Laplace). Looking at the $\k$-space
2P Stokes Ewald sum, and its constituent expressions,
\eqref{eq:ewald_sum_fd}-\eqref{eq:K12}, it is not hard appreciate
the challenge in finding transforms and approximations that mimic the
3P approach. In the electrostatics setting, a variety of non-Ewald
(cf. Mazars \cite{Mazars2005}) and non-2P Ewald methods (cf. Arnold
et al. \cite{Arnold2002} and the references therein), have been
developed instead. These methods have not been adapted to Stokes
(i.e. deriving corresponding \emph{Lekner sums} \cite{Mazars2005} or
correction terms \cite{Arnold2002}), and it is quite evident that, at
best, a substantial effort would be required to do so.

What we propose, in contrast, starts from the mixed sum/integral
\eqref{eq:ewald_sum_fd_int}. Formulating a FFT-based (PME) method
becomes a quite straight-forward matter. It follows the same line of
reasoning that we set forth in \cite{Lindbo2011a}. First, however, we
give a few remarks regarding item $(a)$.

\subsection{Real-space summation in linear time} 
\label{sec:fast_rs_meth}

The truncated real-space sum
\begin{align*}
  \U_{r_c}^R(\x_m) &= \sumn \sum_{\p\in \Z^2}^* \mathbf{1}_{r_c}(\|\x_m
  - \x_n + \tau(\p) \|)A(\xi, \x_m - \x_n + L\p)\f_n,
\end{align*}
with
\begin{align*}
  A(\xi, \x) = 2\left( \frac{\xi e^{-\xi^2 r^2}}{\sqrt{\pi} r^2} +
    \frac{\erfc{(\xi r)}}{2 r^3} \right) (r^2 \id + \x\otimes\x) -
  \frac{4\xi}{\sqrt{\pi}} e^{-\xi^2 r^2}\id,
\end{align*}
is, in order to benefit from much related work, most conveniently
written as
\begin{align}
  \U_{r_c}^R(\x_m) = \sum_{\y\in \NL_m} \hspace{-5pt}A(\xi, \x_m -
  \y)\f(\y), \label{eq:ewald_sum_rs_nl}
\end{align}
where $\NL_m = \NL_m(r_c)$ denotes the set of near neighbors to $\x_m$
(counting periodic images, if necessary). The point being that if
$|\NL_m|$ is constant, for all $m$, as $N$, the number of sources,
grows, then evaluating \eqref{eq:ewald_sum_rs_nl} for all $\x_m$ has
complexity $\O(N)$ instead of $\O(N^2)$. This presupposes that all
neighbor lists $\NL_m$ can be found in linear time, which is the
case. In fact, it is quite elementary and we refer the reader to the
textbook by Frenkel \& Smit \cite[Appendix F]{Frenkel2001}, and the
recent paper \cite{Lindbo2011} by the present authors where additional
details are given. 

Note that each neighbor list, $\NL_m$, can be viewed as the non-zero
pattern for a row in a sparse matrix. In light of this, it's natural to
suggest that the evaluation of \eqref{eq:ewald_sum_rs_nl} for all
$\x_m$ be implemented as a sparse matrix-vector product, $\U^R
\leftarrow \tilde{A}(r_c,\{\x_m\}) \text{vec}(\f)$. The $3N\times 3N$
matrix $\tilde{A}$ has a $3\times3$ block structure corresponding to
Cartesian components $A_{ij}$, but each block has the same sparsity
pattern. If \eqref{eq:ewald_sum_rs_nl} is to be evaluated for many
$\f$, as is the case in the context of iterative solution of boundary
integral equations (cf. Section \ref{sec:bie}), the matrix form saves
a very large amount of redundant arithmetic \cite{Saintillan2005}. In
a setting where locations $\x_m$ change, as in fiber simulations
\cite{Tornberg2006}, the matrix elements of $\tilde{A}$ have to be
recomputed, but the sparsity pattern can be valid for several time
steps\footnote{Typically, the neighbor lists $\NL_m$ are constructed
  with some margin, $\|\x-\y\|\leq r_c + r'$, to allow particles to
  move a distance $r'$ before the neighbor list has to be
  recomputed.}.

\subsection{SE2P Stokes: Fast k-space method} \label{sec:fast_k_meth}

The fast method that we give here is based on earlier work by the
present authors \cite{Lindbo2010,Lindbo2011,Lindbo2011a}, and this
exposition is somewhat condensed and, in contrast to our previous
work, draws in its structure and notation from the elegant treatment of
the 3P electrostatics case by Shan et al. \cite{Shan2005}.

\begin{figure}
  \centering
  \includegraphics{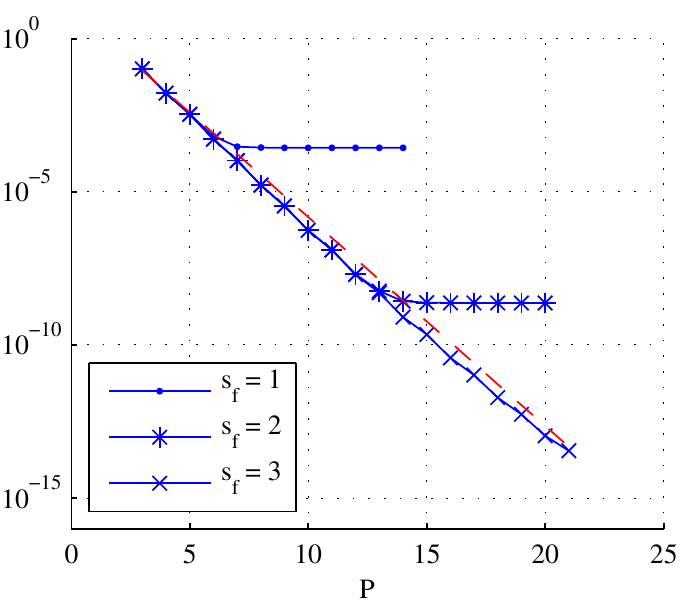}
  \includegraphics{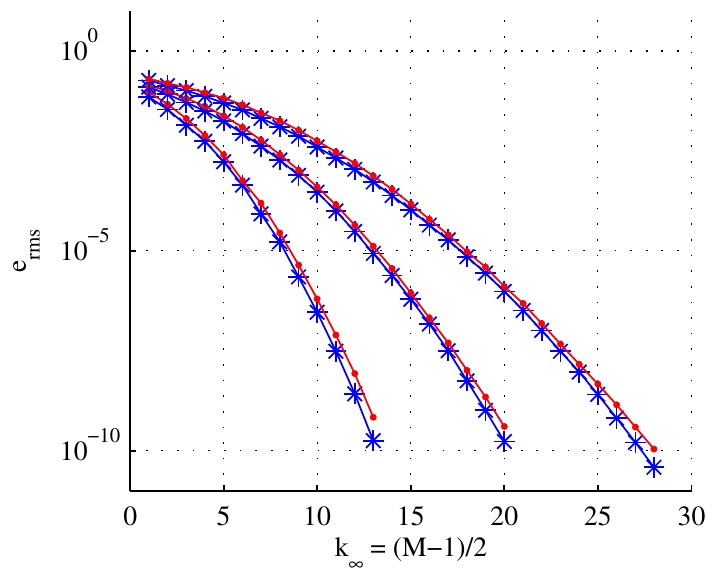}
  \caption{Left: Approximation error in $\infty$-norm for SE2P-Stokes
    method, and (dashed) error estimate \eqref{eq:approx_err_est}, as
    a function of the support of truncated Gaussians, $P$. Here,
    $N=20$, $M=20$, $\xi=4$. Right: ($*$) truncation error for the
    $\k$-space Ewald sum \eqref{eq:ewald_sum_fd}, vs number of modes
    $\kappa_\infty$ (i.e. $|\k|\leq 2\pi\kappa_\infty/L$). As
    ($\cdot$) error in SE2P-Stokes method vs. grid size, where $P$
    selected to make approximation errors small (left). This isolates
    the ``spectrum truncation'' error of the fast method and shows
    that grid size in the fast method can be selected based error
    estimates pertaining only to the underlying Ewald sum.}
  \label{fig:num_res_I}
\end{figure}

Instead of starting from the 2P $\k$-space Stokes Ewald
\eqref{eq:ewald_sum_fd}, we invoke the mixed sum-integral form
\eqref{eq:ewald_sum_fd_int}
\begin{align*}
  \U^F(\r_m,z_m) = \frac{4}{L^2}\sum_{\k\neq 0} \int_\R B(\kf)
  e^{-\nkf^2/4\xi^2} \sumn \f_n e^{i\k\cdot (\r-\r_n)}
  e^{i\kappa(z-z_n)} \d\kappa.
\end{align*}
It is fairly easy \cite{Lindbo2011a} to show that $\U^F(\x_m)$ can be
obtained as a convolution of a particular smooth function,
$\psi:\Omega\times\R\rightarrow\R^3$, with a suitably scaled Gaussian
centered in $\x_m$,
\begin{align}
  \U^F(\r_m,z_m) = C'\int_\R \int_\Omega \psi(\r,z) e^{-\beta \|
    \r-\r_m\|_*^2} e^{-\beta |z-z_m|^2} \d\r \d
  z, \label{eq:se2p_int}
\end{align}
where
\begin{align*}
  C' = \left( \frac{2\xi^2}{\pi\eta} \right)^{3/2}, \quad
  \beta=\frac{2\xi^2}{\eta}
\end{align*}
and $\eta>0$ is a parameter. To define $\psi$, and show how it enters,
we recall the decomposition \eqref{eq:u_decomp} and note that we can
view $B(\kf)e^{-\nkf^2/4\xi^2}$ as the $\k$-space representation of
the ``regularized'' Stokeslet $(S*\gamma)(\x)$, as discussed at the
outset of Section \ref{sec:stokes_ewald_2p_deriv}. In Section
\ref{sec:2p_direc_fd} we proceeded to obtain the Ewald sum
\eqref{eq:ewald_sum_fd} by convolving $S*\gamma$ with $\f(\x)$ and
laboring over the resulting integrals (the convolution itself was
trivial though). Mesh-based Ewald methods deviate at this point. The
logic is to consider not $\f$ but a regularization, $\f_\eta(\x) =
(G(\eta)*\f)(\x)$, and adjust the Green's function accordingly. That
is, choose a kernel $G$ and determine a modified Green's function,
$\tilde{B}$, such that
\begin{align}
  \U_\gamma = (S*\gamma(\xi)*\f)(\x) =
  (\f*G(\eta)*\tilde{B}*G(\eta))(\x). \label{eq:se2p_convs}
\end{align}
It turns out to be advantageous \cite{Lindbo2011} to let $G=C'
e^{-\beta\|\x\|^2}$, so that,
\begin{align}
  \f_\eta(\r,z) = C'\sumn e^{-\beta \| \r-\r_m\|_*^2} e^{-\beta
    |z-z_m|^2} \f_n, \label{eq:se2p_sum}
\end{align}
by which it modified Green's function,
$\tilde{B}=e^{-(1-\eta)\nkf^2/4\xi^2} B(\k,\kappa)$, follows. The
convolution $\f_\eta * \tilde{B}$ is conveniently computed as a
multiplication in frequency domain,
\begin{align}
  \widehat{\psi}(\k, \kappa) = 
  \begin{cases}
    0, & \k=0\\
    e^{-(1-\eta)(k^2+\kappa^2)/4\xi^2} B(\k,\kappa)
    \widehat{\f_\eta}(\k,\kappa), & \mathrm{otherwise}
  \end{cases}. \label{eq:se2p_mul}
\end{align}
The last convolution in \eqref{eq:se2p_convs}, $(\psi*G(\eta))(\x)$,
brings us back to \eqref{eq:se2p_int}.  In
\cite{Lindbo2010,Lindbo2011,Lindbo2011a} detailed and constructive
derivations are given instead of this sketch. To clarify, we give a
few remarks, starting with a summary of the method:
\begin{description}
\item[Summary of method] ~\\ To compute the (non-singular) $\k$-space
  contribution $\U^F(\x_m)$ for all $m$ the steps are as follows:
  $(i)$ evaluate $\f_\eta$ \eqref{eq:se2p_sum} on a uniform grid over
  $\Omega$; $(ii)$ compute a mixed Fourier transform to arrive at
  $\widehat{\f_\eta}$; $(iii)$ multiply with modified Green's function
  according to \eqref{eq:se2p_mul}; $(iv)$ an inverse mixed transform
  yields $\psi$ on a regular grid, so that; $(v)$ the convolution
  \eqref{eq:se2p_int} can be computed for all $\x_m$. In steps $(i)$
  and $(v)$, the Gaussians are truncated to have support on $P^3$ grid
  points, for some $P$ determined by the acceptable approximation
  error.
\item[Fast transforms; uniform grids] PME-methods become efficient by
  requiring that the regularized charge distribution $\f_\eta(\x)$ be
  evaluated on a uniform grid. The transforms are then handled by
  FFTs.
\item[Mixed transforms] Recall the spectral representation
  \eqref{eq:mixed_ft} of 2P functions in terms of discrete Fourier
  series in $(x,y)$ and a continuous transform in $z$. Hence, the
  transforms $\f_\eta \rightarrow \widehat{\f_\eta}$ and
  $\widehat{\psi}\rightarrow \psi$ are mixed. This, together with the
  exclusion of $\k=0$ in \eqref{eq:se2p_mul} and the inclusion of the
  singularity contribution \eqref{eq:ewald_sum_k0}, is where our
  proposed 2P method deviates from the well-established 3P methods.
\item[Fourier integral quadrature] Computing the mixed transforms
  requires two standard discrete Fourier transforms and an
  approximation of the Fourier integral (in the $z$-direction). The
  quadrature step has to be done in the same time-complexity as the
  discrete transforms; otherwise, the method would not be more
  efficient than directly summing \eqref{eq:ewald_sum_fd}. This is
  discussed at length in \cite{Lindbo2011a}, and it is shown to be
  quite satisfactory to use an FFT-based method (following Press
  et al. \cite[Sec.13.9]{Press2007}) on a moderately oversampled
  grid.
\item[Spectral accuracy] Trapezoidal quadrature applied to
  \eqref{eq:se2p_int} is spectrally accurate, due to the
  $C^\infty$-regularity of the integrand. Detailed analysis on this
  appears in \cite{Lindbo2011}. Moreover, the accuracy in computing
  the Fourier integral in the mixed transform using a simple FFT-based
  approach strongly depends on the regularity of the integrand.
\item[Grid size corresponds to direct sum truncation] As is also
  extensively discussed in \cite{Lindbo2011}, the proposed method
  enjoys a close relationship between the Ewald sum
  \eqref{eq:ewald_sum_fd} and the fast method of this section. A
  finite truncation $|\k|\leq2 \pi \kmax/L$ of the Ewald sum
  \eqref{eq:ewald_sum_fd} corresponds to a grid of size $M=2
  \kmax+1$. Due to the construction of the PME-method based on
  Gaussians (with $C^\infty$-regularity), the truncation error from
  the Ewald sum carries over to the fast method, meaning that the
  appropriate grid size can be estimated using truncation error
  estimates for the Ewald sum. This is, as of yet, conjecture in the
  Stokes 2P setting, verified numerically (cf. Figure
  \ref{fig:num_res_I}, right).
\item[Parameters and truncation] The free parameter, $\eta$, can be
  used to control the width, denoted $w$, of the Gaussian used for the
  convolutions in \eqref{eq:se2p_sum}, \eqref{eq:se2p_int}, and we
  find it natural to prescribe this width in terms of a number of grid
  points, $w=LP/(2M)=hP/2$. Gaussians lack compact support, but they
  are highly localized. It is natural to truncate them, as is done in
  the non-uniform FFT \cite{Greengard2004a,Dutt1993}. We let $P \leq
  M$ denote the number of grid points within the support of each
  Gaussian, as seen in Figure \ref{fig:gaussian} (bottom). Moreover,
  it's important to have control of the shape, parameterized with
  $m>0$, of the Gaussians (independently of $\xi$), see Figure
  \ref{fig:gaussian} (top). It then follows naturally
  \cite{Lindbo2011} to let
  \begin{align}
    \eta = \left(\frac{2w\xi}{m}\right)^2. \label{eq:eta_general}
  \end{align}
  \begin{figure}
    \centering
    \includegraphics{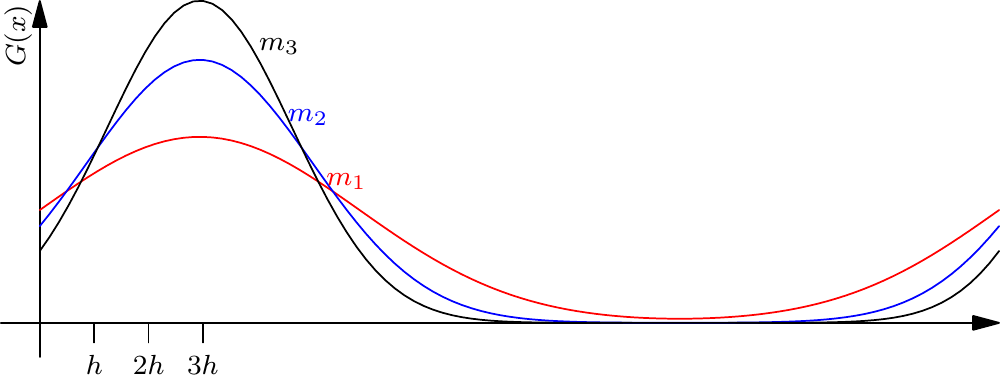}
    \includegraphics{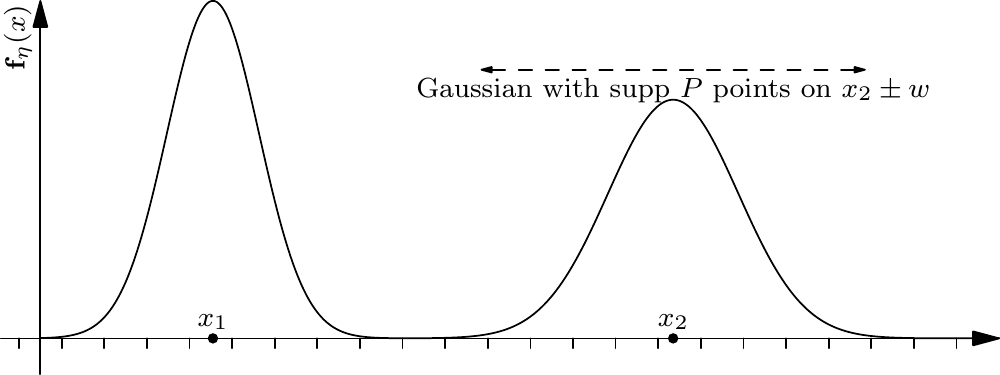}
    \caption{Top: Gaussians with different shape parameters,
      $m_1<m_2<m_3$. Bottom: Gaussian with support on $P$ grid points
      around $\x_j$}
    \label{fig:gaussian}
  \end{figure}
\item[Periodicity] Above, $\|\cdot\|_*$ denotes that periodicity is
  implied. To prove that \eqref{eq:se2p_int} equals
  \eqref{eq:ewald_sum_fd_int} one needs this notion to be exact (which
  rules out the well-known ``closest image convention''). Formally, as
  Gaussians lack compact support, it should be understood e.g. that
  $e^{-\beta\|\x-\x_m\|^2_*}:= \sum_{\p\in\Z^2}e^{-\beta\|\x-\x_m +
    \tau(\p)\|^2}$. After truncating the Gaussians, this ceases to be
  relevant -- simply extend the domain to accommodate the support of
  the truncated Gaussians; evaluate \eqref{eq:se2p_sum} on this
  domain, taking Euclidean distance between particles and grid points;
  and then additively wrap the extended domain into the original one.
\item[Approximation errors] By approximation errors we mean the errors
  that the fast SE2P method contributes (in addition to the
  \emph{spectrum truncation} errors that are inherited from a finite
  truncation of \eqref{eq:ewald_sum_fd}).  These fall into three
  categories: $(i)$ the numerical error from evaluating the integral
  \eqref{eq:se2p_int} with a trapezoidal rule $T_P$; $(ii)$ the
  truncation of the Gaussians; and, finally, $(iii)$ the error in
  computing the Fourier integral in the mixed transforms. Regarding
  $(i)$ and $(ii)$, we can prove a detailed error estimate
  \cite[Thm. 3.1]{Lindbo2011}, namely that
  \begin{align}
    | T_P - \U^F | \leq C\left( e^{-\pi^2 P^2/(2 m^2)} +
      \erfc\left(m/\sqrt{2}\right)\right). \label{eq:approx_err_est}
  \end{align}
  Naturally the first term corresponds to $(i)$ and the second to
  $(ii)$. It also reveals that if $m$ is taken large the first term
  goes slower to zero, indicating that selecting $m\approx\sqrt{\pi
    P}$ strikes a balance between the terms that is favorable for the
  total error. The final, and 2P specific, error concerns the Fourier
  integral quadrature, and is, as noted above, examined in detail in
  \cite{Lindbo2011a}.
\item[Fast gridding] The main trade-off for a spectral method is that
  Gaussians have wider support than e.g. the Cardinal B-splines (used
  in the SPME method \cite{Essmann1995}) have. By using the \emph{fast
    Gaussian gridding} (FGG) procedure, proposed by Greengard \& Lee
  for the non-uniform FFT \cite{Greengard2004a}, we mediate this to a
  large extent. Again, this is examined in \cite{Lindbo2011}, and
  detailed algorithms are given.
\end{description} 

\subsection{The singular term} \label{sec:fast_k0_meth}

The last term to evaluate is the contribution from the singular
integrals that were taken out of the $\k$-space sum
\eqref{eq:ewald_sum_fd},
\begin{align}
  \Ufz(z_m) = -\frac{4}{L^2} \sumn g(z_m-z_n)
  \id_2\f_n. \nonumber\\
  g(z) := \pi z \erf\big( z \xi \big) +
  \frac{\sqrt{\pi}}{2\xi} e^{-z^2 \xi^2}, \label{eq:k0_kernel}
\end{align}
cf. the Laplace case \cite{Lindbo2011a}. Because $g$ only depends on
$z$ it's almost trivial to compute $\Ufz$ using an interpolation
approach. We advocate a Chebyshev method as follows: $(i)$ evaluate
$\Ufz(s)$, where $s$ denotes the set of $M_0$ Gauss-points scaled to
the relevant interval; $(ii)$ compute the $M_0$ coefficients of the
interpolating Chebyshev polynomial, $\zeta(z)$, by means of elementary
recursions; $(iii)$ evaluate $\Ufz(z_m) \approx \zeta(z_m)$ using a
numerically stable Clenshaw formula
\cite[Sec. 5.4]{Press2007}. Properties of Chebyshev interpolation,
including spectral accuracy, are discussed in e.g. Rivlin
\cite{Rivlin1990} and many elementary textbooks.

\subsection{Parameters, estimates and modus operandi} 
\label{sec:param_selection}

A commonly cited drawback for Ewald methods, particularly for fast
FFT-based ones, is that there are several parameters to choose
appropriate values for. For 3P Laplace, this is extensively discussed
by Deserno \& Holm in their famous survey \cite{Deserno1998}. Recent
results and extensive discussion is found in \cite{Wang2010}. In the
2P Laplace case, the situation is, unsurprisingly, quite misty. Kawata
\& Nagashima \cite{Kawata2001b} have proposed a method, which shares
some foundations with the present work, where it's suggest that an
optimization approach be applied to tune a set of eight
parameters. The situation is, hopefully, clarified to a large extent
in \cite{Lindbo2011a}.

We suggest that the parameters fall into three natural categories:
\begin{enumerate}
\item The desired accuracy, $\epsi$, in the computed $\U$ with respect
  to a particular norm.
\item The Ewald decomposition parameter, $\xi$, and the truncation of
  real- and $\k$-space Ewald sums, $r_c$ and $\kmax$, for particular
  $\epsi$.
\item Parameters relating to the fast method:
  \begin{itemize} 
  \item[a)] the grid size, $M$, (in each dimension);
  \item[b)] the support, $P$, and shape, $m$, of Gaussians, in the
    FFT-based method for $\U^F$, and the oversampling factor $s_f$ in
    the associated mixed transforms;
  \item[c)] the number of Gauss-points, $M_0$, in the
    interpolation method for $\Ufz$.
  \end{itemize}
\end{enumerate}

Item one, choosing $\epsi$, comes first -- all other parameters will
depend on it. We noted that the cost of the real-space sum can vary by
orders of magnitude depending on it's implementation and whether a
sparse matrix can be assembled and reused or not. With this in mind,
we argue that the next consideration should be to select the
real-space truncation $r_c$ such that the real-space is cheap to
compute.

Item two is clarified by the truncation error estimates of Section
\ref{sec:trunc_est}. We noted that, in terms of $\epsi$ and $r_c$ we
can take
\begin{align}
  \xi = \frac{1}{r_c} \sqrt{\log
    \left(\frac{2}{\epsi}\sqrt{\frac{Q r_c}{L^3}}\right)}.
\end{align}
That is, $\xi$ is chosen large enough that the error committed by
truncating the real-space sum at $r_c$ is close to $\epsi$. Fourth, we
invoke the equivalence between $\k$-space truncation and grid in the
PME-type method. That is, the truncation error estimate
\eqref{eq:err_F_rms_heuristic} is inverted,
\begin{align}
  \kmax=\frac{\sqrt{3}L\xi}{2 \pi }\sqrt{W\left(\frac{4 L^{2/3} \xi ^2
        Q^{2/3}}{3 \pi ^{10/3} \epsi ^{4/3}}\right)},
\end{align}
where $W(\cdot)$ is Lambert W function \cite{Corless1996}, and the
grid size selected as $M=2\kmax+1$.

The remaining parameters, $\{P,m,s_f\}$, are all accuracy parameters,
pertaining to the approximation errors added by the fast method, that
don't depend on the system. The one with greatest impact on the
computational cost is $P$, the discrete support of the truncated
Gaussians, and (cf. Figure \ref{fig:num_res_I}, left) rough estimates
are that $P=15$ is appropriate for accuracy around $10^{-10}$ and
$P=23$ is required for full (double-precision) accuracy. As previously
determined, it's natural to let $m=C\sqrt{\pi P}$, and we've found
that this constant is best taken slightly below unity, $C=0.92$. The
FFT-oversampling in the $z$-direction, $s_f$, should be at least two,
but should be four or six for higher accuracies (cf. Figure
\ref{fig:num_res_I}). Finally, the number of Gauss-points in the 1D
interpolation used to compute $\Ufz$ should be small, $M_0<100$,
typically around 50.

We would like to emphasize that, although there are a number of
parameters to select, the procedure is eminently straight-forward. In
particular, PME-grid size selection based on a truncation error
estimate for the underlying Ewald sum, is an important feature.

\section{Numerical results} 
\label{sec:num_res}

In the introduction several areas of ongoing research involving
boundary integral methods for Stokes were noted, and we would like to
put the present method into that context. This also gives the
opportunity to illustrate parameter selection and the practical
characteristics of our method.

\subsection{Fast boundary integral methods in a periodic setting}
\label{sec:bie}

As an example, consider a large number of rigid spheres, $\Gamma_i,
i=1\dots N_s$, in 2P Stokes flow. Let $\tf_i$ denote the (unknown)
Stokeslet density on $\Gamma_i$. Under the constraint of planar
periodicity we write the velocity field generated by all $\{\tf_i,
\Gamma_i\}$ as
\begin{align}
  \U(\x) = \frac{1}{8 \pi \mu }\sump{2} \sum_{i=1}^{N_s}
  \int_{\Gamma_i} S(\x-\y+\tau(\p)) \tf_i(\y) \d\Gamma(\y). 
  \label{eq:u_spheres}
\end{align}
Recall that the periodized Stokeslet sum \eqref{eq:stokeslet_sum} was
given a well-defined meaning by the 2P Stokeslet Ewald decomposition
(Section \ref{sec:stokes_ewald_2p}). The same logic shall apply to
\eqref{eq:u_spheres}. We refer the reader to
e.g. \cite{Pozrikidis1992,Ying2006} regarding which Green's functions
are required to represent particular flow cases, noting that for this
external flow it suffices to use the Stokeslet.

The boundary integral setting is vastly expanded in comparison to the
exposition hitherto. Issues that arise, as concisely reviewed in Ying
et al. \cite{Ying2006}, include fast summation methods (to which this
work pertains), accurate quadrature rules (surveyed below) and domain
boundary representation.

Let each sphere $\Gamma^i$ be discretized with a set of quadrature
points, $\x^i_j$, $j=1,\dots,N_p$, and let $T_\x$ denote integration with
respect to $\x$ by a simple quadrature rule, $T[\zeta] := \sum_j
\zeta(\x_j) w_j \approx \int_\Gamma \zeta(\x) \d
\Gamma(\x)$. Moreover, let $T^{0,k}$ denote the so-called
``punctured'' rule, where $w_k\equiv 0$.

For $\x$ on some $\Gamma_i$, the integral over $\Gamma_i$ in
\eqref{eq:u_spheres} is singular when $\p=0$. Discretizing it in the
Nystr\"{o}m fashion, we apply the punctured rule,
\begin{align*}
  \U^i(\x_m) &\sim \int_{\Gamma_i} S(\x_m-\y) \tf(\y) \d \Gamma(\y)\\
  & \approx T_\y^{0,m}[S(\x_m-\y) \tf(\y)].
\end{align*}
This has one major benefit and one drawback. The benefit is that it
renders the discretization of \eqref{eq:u_spheres} equivalent to the
2P Stokeslet ewald sum \eqref{eq:stokeslet_ewald_sum_2p} -- c.f. the
exclusion of the term $\{n=m, \p=0\}$ in the real space sum
\eqref{eq:ewald_sum_rs}. The drawback is that the punctured rule is
only first order accurate.

To obtain higher accuracy, without sacrificing the structure that lets
us apply the Ewald decomposition, one can add a local correction,
$\quadcorr^m$, to the punctured rule,
\begin{align*}
  \U^i(\x_m) \approx T_\y^{0,m}[S(\x_m-\y) \tf(\y)] + \quadcorr^m.
\end{align*}

Such local corrections to simple quadrature rules (that allow
integrable singularities to be handled) have been extensively
investigated, and are, of course, unrelated to the planar periodicity
of the present problem. Progress has been on a case-by-case basis,
with different corrections developed depending on the singularity and
the geometry of the boundary, $\Gamma$. Early references here include
Lyness \cite{Lyness1976} and Kapur \& Rokhlin \cite{Kapur1997}, with
subsequent work by e.g. Aguilar \& Chen \cite{Aguilar2005} and Marin
et al. \cite{Marin2011}. The latter gives formal proofs to the
effect, roughly, that on a uniform grid in $d$ dimensions where the
singularity has order $q$ and $p$ is the radius (in terms of grid
points) of the modified quadrature rule, accuracy of order
$\O(h^{2p+2+q+d})$ is attained. Furthermore, Marin et al. in
\cite{Marin2011a} developed such formulas for the Stokeslet, $S$,
though only of the case of $\Gamma$ being flat.

An alternative that could be expected to yield local corrections, in a
way suitable for pairing with fast Ewald methods, is the contour
integral formulation of Bazhlekov, Anderson and Meijer
\cite{Bazhlekov2004}. Other alternatives for quadrature over
integrable singularities includes \emph{singularity subtraction}, as
discussed in the rich survey by Pozrikidis \cite{Pozrikidis2001}, and
methods based on carefully selected variable transformations, as
discussed in e.g. Sidi \cite{Sidi2006,Sidi2008} and Ying
et al. \cite{Ying2006}.

Moving on, for $\x_m$ on $\Gamma_i$, we discretize
\eqref{eq:u_spheres} as
\begin{align}
  \begin{split}
    \U(\x_m)\approx \frac{1}{8\pi\mu} \left(T^{0,m}_\y[S(\x_m -
      \y)\tilde{\f^i}(\y)] +\quadcorr^m+
      \sum_{j=1}^{N_s}\sum_{\p\in\Z^2\setminus 0}
      T_\y[S(\x_m-\y+\tau(\p)) \tilde{\f^j}(\y) ]
    \right)\\
    = \frac{1}{8\pi\mu} ( \Ewp(\x, \f) + \quadcorr^m(S,\f) ).
  \end{split} \label{eq:discr_BIE}
\end{align}
Here, $\x$ denotes the set of $N = N_s N_p$ discretization points
(i.e. $\x = \{\x^i_j: i=1,\dots,N_s, j=1,\dots,N_p
\}$). Correspondingly, $\f$ denotes the set of discrete Stokeslet
strengths multiplied by appropriate quadrature weights,
$\f=\{\tilde{\f}^i_j w_j \}$.

The mapping $\f \rightarrow \U$, for fixed $\x$, is to be understood
as a linear system of dimension $3N \times 3N$; knowing $\U$, we can
solve for $\f$. Such an inversion is by necessity iterative, as the
mapping does not have a closed form. Indeed, the Ewald summation
method is required to make the action of the periodized Stokeslet
convergent (including constraints on $\f$ and physical conditions at
$z\rightarrow\pm\infty$).

The fast Ewald method of Section \ref{sec:fast_method} acts as an
$\O(N\log N)$ ``matvec'' operator, for use inside an iterative solver
(such as GMRES). The ultimate complexity then depends on the
convergence of this iterative procedure (whether the number of GMRES
iterations depends on $N$ or not), which leads to the expansive topic
of preconditioning; Ying et al. \cite{Ying2006} use a method due to
Greengard et al. \cite{Greengard1996}, wherein additional references
are found.

In this context it's natural to remark that a second-kind boundary
integral formulation is expected to be more advantageous with respect
to iteration convergence than the first-kind equation
\eqref{eq:u_spheres}. This motivates the development of Ewald
decompositions, and associated fast methods, for the double-layer
(``doublet'' or ``stresslet'') Stokes potential.

Finally, it should be noted that investigations of physical systems
generally pose the boundary integral problem in a different
form. Rather than having boundary conditions for $\U$ on all
particles, kinematic conditions (which include boundary integration)
are used and the system is closed using force and torque
balances. Such a formulation is used for sedimenting fibers
\cite{Tornberg2004, Saintillan2005, Tornberg2006}. Ewald techniques
apply and can handle the main computational task in these
sedimentation problems, though additional considerations emerge.

\subsection{Stokeslet Ewald as a matvec operator}

Having posed a relevant boundary integral problem, we now discuss how
to deploy our method in detail and give an indication of how efficient
it is. Several of the topics touched upon in the previous section,
such as preconditioning and local quadrature correction terms, are
beyond the scope of the present work. A central concern is clarifying
how to balance the computational cost of the terms in
\eqref{eq:stokeslet_ewald_sum_2p}.

\subsubsection{Parameters, accuracy and cost}

\begin{figure}
  \centering
  \includegraphics{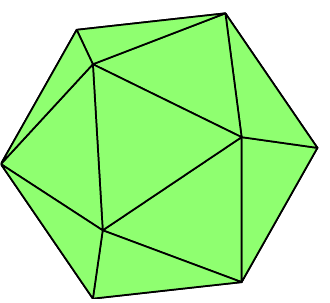} 
  \includegraphics{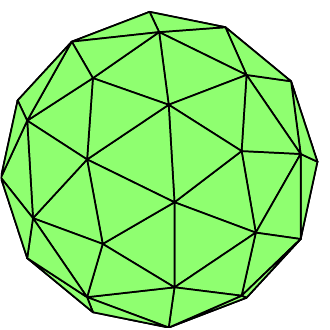}
  \includegraphics{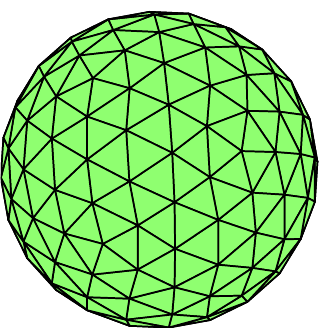}
  \caption{Nearly uniform point distribution on sphere by subdividing
    icosahedron. Left to right: $N_p=12,42,162$.}
  \label{fig:spheres}
\end{figure}

We consider a system with $N_s$ spheres, each with a radius $r_s$, in
a domain $\Omega=[0, L)^3$, and assume that the spheres' center are
separated by at least $3r_s$. A nearly uniform distribution of points
on each sphere, as seen in Figure \ref{fig:spheres}, is obtained by
subdividing an icosahedron \cite{Luigi2009}.

We target an accuracy $\epsi \approx 10^{-9}$ in the Ewald method. The
cost of computing the 2P Stokeslet Ewald sums \eqref{eq:ewald_sum_rs}
and \eqref{eq:ewald_sum_fd} depends on the density of the
system. Throughout, we shall take an initial number of spheres $N_s^0$
in a domain with $L=L^0$ and then grow the system at constant density
to, say, $L=3L^0$.

Assuming uniformity, the number of points within a ball with radius
$r_c$ is $N_n(r_c) = 4\pi r_c^3 N_s N_p/(3 L^3)$. This represents the
number of near neighbors that have to be considered for a particular
real-space truncation radius, $r_c$. The cost of computing the
real-space sum (cf. Section \ref{sec:fast_rs_meth}) is roughly $N_s
N_p N_n(r_c) \lambda$, where $\lambda$ represents the arithmetic cost
associated with each interaction, plus the cost of finding the
neighbor list. The factor $\lambda$ is one (i.e. one floating-point
multiply-add) if the real-space sum is computed as a sparse
matrix-vector multiplication and on the order of 100-1000 otherwise
(because of the computations of $\erf(\cdot)$ and $\exp(\cdot)$).

Table \ref{tab:param} attempts to give an overview of the relationship
between real-space sum cost, the system size, and what is then implied
for the $\k$-space sum by the inverted error estimates of Section
\ref{sec:param_selection}. As expected, making the systems denser
requires $\xi$ to grow for the cost of the real-space sum (in terms of
near neighbors) to remain fixed. Correspondingly, the FFT-grid
$M=2\kmax$ grows to maintain a fixed accuracy. In what remains, we
shall take $P=16$, the FFTs oversampled by a factor four in the
$z$-direction and $M_0=80$ Gauss points in interpolation method for
\eqref{eq:ewald_sum_k0}. With $r_c = 1/2, \xi=9, M=30$, and $\x$ the
points from 5 randomly placed spheres, we compute RMS errors:
$2.2\times 10^{-9}$ and $6.2 \times 10^{-9}$ for the real-space sum
and SE2P method respectively.

\begin{table}
  \centering
  \begin{tabular}{l|l}
    Given & Derived \\
    \hline
    $N_s=100, N_p=12, N_n=100$ & 	 $r_c=0.27, \xi=17.7, \kmax=27 $ \\ 
    $N_s=100, N_p=12, N_n=1000$ & 	 $r_c=0.58, \xi=8.3, \kmax=12 $ \\ 
    \\
    $N_s=100, N_p=42, N_n=100$ & 	 $r_c=0.18, \xi=26.8, \kmax=41 $ \\ 
    $N_s=100, N_p=42, N_n=1000$ & 	 $r_c=0.38, \xi=12.5, \kmax=19 $ \\ 
    \\
    $N_s=1000, N_p=12, N_n=100$ & 	 $r_c=0.13, \xi=37.9, \kmax=58 $ \\ 
    $N_s=1000, N_p=12, N_n=1000$ & 	 $r_c=0.27, \xi=17.7, \kmax=27 $ \\ 
    \\
    $N_s=1000, N_p=42, N_n=100$ & 	 $r_c=0.08, \xi=57.2, \kmax=89 $ \\ 
    $N_s=1000, N_p=42, N_n=1000$ & 	 $r_c=0.18, \xi=26.8, \kmax=41 $ \\ 
  \end{tabular}
  \caption{Parameter examples for 2P Stokeslet Ewald summation, with 
    $\epsi=10^{-10}$ 
    and $L=1$. $N_s$: number of spheres. $N_p$: number of points per sphere. 
    $N_n$: number of near neighbors to account for in real-space sum. Right 
    column: parameter values selected based on truncation error estimates 
    \eqref{eq:err_R_rms_simpl} and \eqref{eq:err_F_rms_heuristic}.}
  \label{tab:param}
\end{table}

\subsubsection{Fast implementations and run-time profiles}

\begin{figure}
  \centering
  \includegraphics{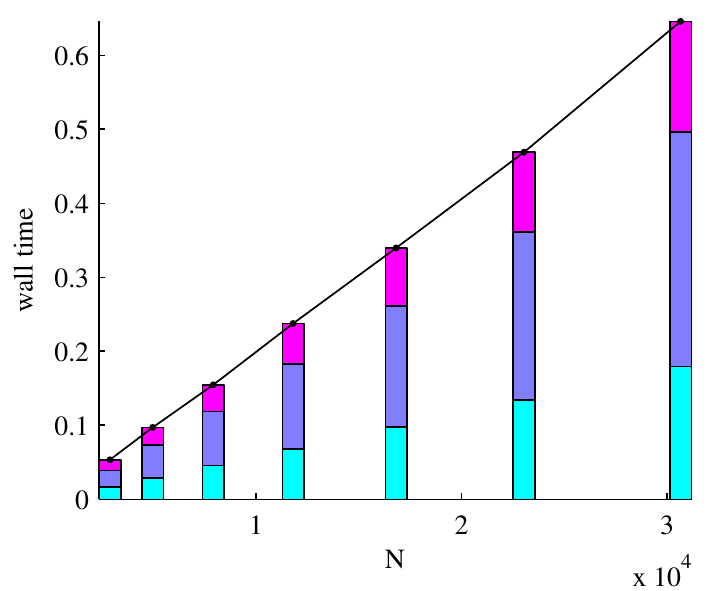}
  \includegraphics{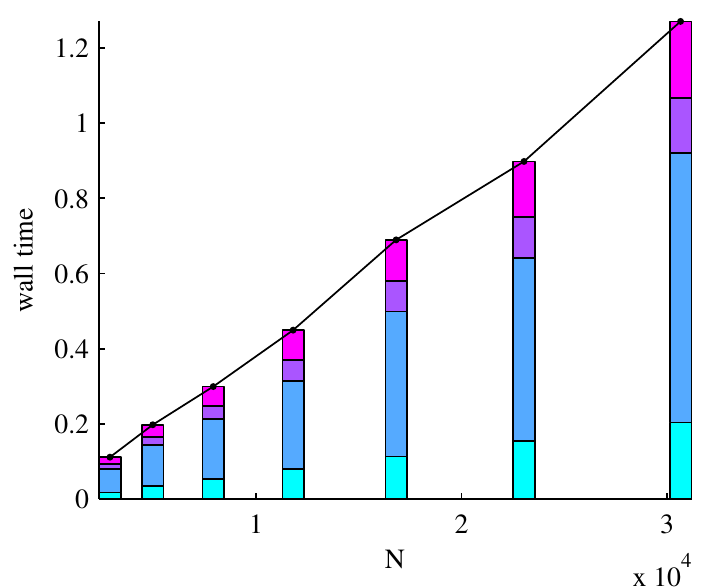}

  ~\\
  $\xi=9$, $r_c=1/2$, $\epsi=10^{-9}$, $N_n\approx 400$\\
  \begin{tabular}{l|lllllll}
    \hline
    $L=$& 1 & 1.2 & 1.4 & 1.6  & 1.8  & 2 & 2.2\\
    $N_s=$ & 80 &  138 &  220 &  328 &  467 &  640 &  852\\
    $N=$& 2880 & 4968 & 7920 & 11808 & 16812 & 23040 & 30672\\
    $M=$ & 30& 36&    42&    48&    54&    60&    66
  \end{tabular}

  \caption{Runtime to compute \eqref{eq:stokeslet_ewald_sum_2p} as
    system is scaled up, keeping the number of near neighbors fixed at
    around 400, according to the table above. {\bf Left:} Real-space
    sum, cf. Remark \ref{rem:rs}. {\bf Right:} FFT-based $\k$-space
    method, cf. Remark \ref{rem:fd}.}
  \label{fig:time_breakdown_I}
\end{figure}

\begin{figure}
  \centering
  \includegraphics{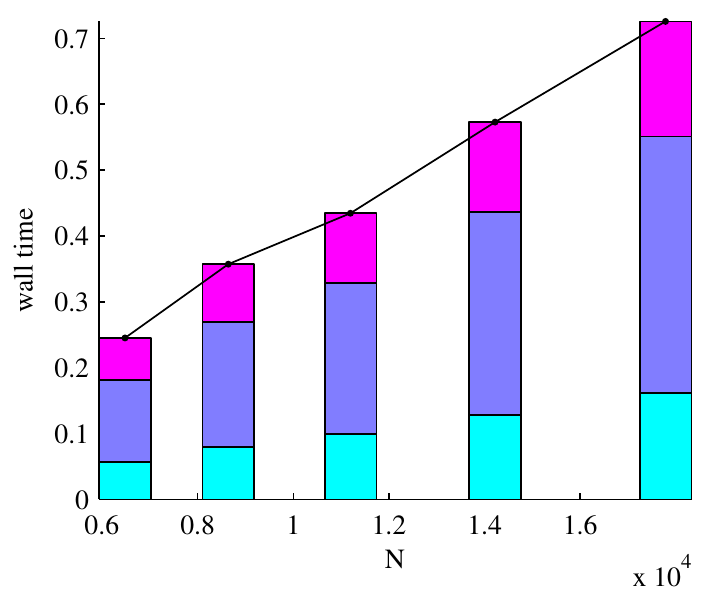}
  \includegraphics{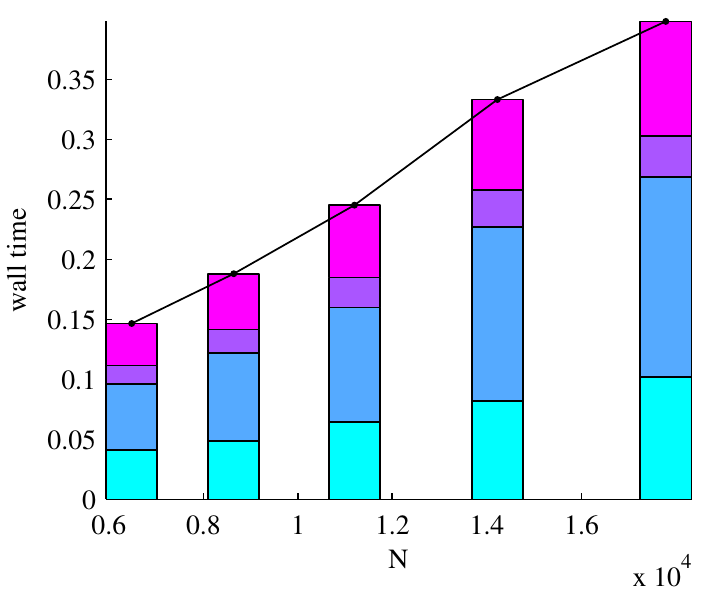}

  ~\\
  $\xi=9$, $r_c=1/2$, $\epsi=10^{-9}$, $N_n\approx 1000$\\
  \begin{tabular}{l|lllllll}
    \hline
    $L=$& 1 & 1.1 & 1.2 & 1.3 & 1.4\\
    $N_s=$ &  180 & 240 & 311 & 395 & 494\\
    $N=$& 6480 & 8640 & 11196 & 14220 & 17784\\
    $M=$ & 30  &  33  &  36  &  39  &  42
  \end{tabular}

  \caption{Runtime to compute \eqref{eq:stokeslet_ewald_sum_2p} as
    system is scaled up, keeping the number of near neighbors fixed at
    around 1000, according to the table above.  {\bf Left:} Real-space
    sum, cf. Remark \ref{rem:rs}. {\bf Right:} FFT-based $\k$-space
    method, cf. Remark \ref{rem:fd}. }
  \label{fig:time_breakdown_II}
\end{figure}

In Figures \ref{fig:time_breakdown_I} and \ref{fig:time_breakdown_II}
we give timing results as the system is scaled up and show how the
run-time is distributed between the steps in the algorithm. Here we
fix the average number of near neighbors at roughly 400 and 1000
respectively, and scale the system up under that constraint. The
following remarks clarify the situation:

\begin{rmrk}[Real space computation] \label{rem:rs} The real space sum
  is evaluated in the matrix form as discussed in Section
  \ref{sec:fast_rs_meth}. The matrix is computed once and the time to
  do so is amortized over a hypothetical number of iterations: 50
  (which is reasonable, see \cite{Saintillan2005}). This
  pre-computation contains two parts: first the matrix elements are
  computed (a), then the sparse matrix is finalized (b). The
  real-space contribution \eqref{eq:ewald_sum_rs} is computed as a
  matrix-vector product (c). The bars in Figures
  \ref{fig:time_breakdown_I} and \ref{fig:time_breakdown_II} (left)
  are stacked from bottom to top in the order (a), (b), (c) -- showing
  that, at 50 iterations, constructing the sparse matrix is still the
  dominating cost.
\end{rmrk}

\begin{rmrk}[Computation of $\k$-space method] \label{rem:fd} The
  tasks in the FFT-based method to compute $\U^F$ are as follows: (a)
  compute grid function \eqref{eq:se2p_sum}; (b) compute oversampled
  transforms; (c) solve Poission problem \eqref{eq:se2p_mul}; and (d)
  compute convolution with Gaussians \eqref{eq:se2p_int}. The bar
  stacks in Figures \ref{fig:time_breakdown_I} and
  \ref{fig:time_breakdown_II} (right) breaks down the total time for
  the method in these steps from bottom to top. 

  Our implementation was mostly in Matlab code, and, hence, FFTs were
  handled by the highly optimized library FFTW
  \cite{Frigo2005}\footnote{By convention, the fast Fourier transform
    is said to have complexity $\O(N\log N)$, but this concept
    encompasses a lot of variability which is seen in practice. For
    instance, an FFT of length $2^n$ will typically be several times
    faster than a transform of length $2^n+1$. The grid sizes seen in
    the scaling tests here try to avoid the best and worst case
    transform lengths, to faithfully represent the method.}. The code
  for gridding \eqref{eq:se2p_sum} and integration \eqref{eq:se2p_int}
  with the FGG algorithm \cite{Greengard2004a, Lindbo2011} was
  implemented in C and the time-critical parts were hand-coded at
  machine level with SSE instructions and software prefetching. These
  implementations run close to peak flops of present hardware. An
  untuned implementation in C runs roughly at a third of the speed,
  making the gridding steps dominate transforms.
\end{rmrk}

\begin{rmrk}[Computation of $\Ufz$] The computation of $\Ufz$ as
  discussed in Section \ref{sec:fast_k0_meth} involves computing $\pi
  z \erf\big( z \xi \big) + \frac{\sqrt{\pi}}{2\xi} e^{-z^2 \xi^2}$
  for $z= z_m - z_n$, where $n=1,\dots,N$ and $m=1,\dots, M_0$ (the
  number of Gauss-points for the Chebyshev interpolation). In the
  iterative setting, where we wish to rapidly evaluate $\Ufz$ for many
  $\f$, this arithmetic should be pre-computed and stored in a matrix,
  $A_0$. Then the sum in \eqref{eq:k0_kernel} is computed by
  multiplying $A_0$ with $\f_i$, $i=1,2$. The interpolation procedure
  follows, which is very fast. Corresponding to the runs in Figures
  \ref{fig:time_breakdown_I} and \ref{fig:time_breakdown_II}, the
  times to compute $\Ufz$ were vanishingly small.
\end{rmrk}

We note that balancing the computational cost of the real-space sum
and $\k$-space fast method is non-trivial. At a basic level, it hinges
on how dense the system is. Evaluating the real-space sum using a
neighbor list method (matrix-form or not) requires the list to be
stored. In the iterative setting, the matrix form offers superior
performance -- so much so that the main constraint of the real-space
method becomes memory (i.e. limiting $r_c$ for a particular $N$). In
the fast $\k$-space method the split in computational burden between
transforms and gridding also depends on density -- or rather, it
depends on the Ewald parameter $\xi$, which has to increase to keep
the number of near neighbors to account for in the real-space sum
fixed if the system is made denser. 

\section{Conclusions}

In this paper we have derived a Ewald decomposition for the Stokeslet
in planar periodicity (Section \ref{sec:stokes_ewald_2p_deriv}) and a
PME-type $\O(N \log N)$ method for the fast evaluation of the
resulting expression. The decomposition is the natural 2P counterpart
to the classical 3P decomposition by Hasimoto
\cite{Hasimoto1959}. Truncation error estimates are provided to aid in
selecting parameters (Section \ref{sec:trunc_est}). The fast,
PME-type, method (Section \ref{sec:fast_method}) for computing the
$\k$-space sum \eqref{eq:ewald_sum_fd} is based on a mixed
sum/integral form \eqref{eq:ewald_sum_fd_int}, and is similar to the
method by the present authors \cite{Lindbo2011a} for the corresponding
problem in electrostatics. This appear to be the first fast method for
computing Stokeslet Ewald sums in planar periodicity, and has three
attractive properties: it is spectrally accurate; it uses the minimal
amount of memory that a gridded Ewald method can use; and a clear view
of numerical errors and how to choose parameters is provided (Section
\ref{sec:param_selection}). The final part explores the practicalities
of the proposed method, and surveys the computational issues involved
in applying it to 2-periodic boundary integral Stokes problems. We
presently pursue applications-oriented questions in this regard, and
hope to communicate further results in the near future.

\appendix
\section{2P Stokeslet sums, details}
\subsection{Explicit $\k$-space form of Stokeslet sum} 
\label{app:non-screened_u_integrals}
Here we compute
\begin{align}
  \tU(\x_m) = \frac{4}{L^2} \sum_{\k\neq 0} \int_\R \left( \frac{\id}{\nkf^2} -
    \frac{\kf \otimes \kf}{\nkf^4}\right) \sumn \f_n e^{i \kf \cdot
    (\x_m - \x_n )}\d \kappa, \label{eq:uf_deltas}
\end{align}
where $\kf = (\k, \kappa) = (k_1, k_2, \kappa)$ is the composition of
the discrete and continuous transform variables. The integrals in
\eqref{eq:uf_deltas} are computable. We let
\begin{align*}
  \tU(\x_m) = \frac{4}{L^2} \sum_{\k\neq 0} \sumn \tQ(\k,z_m-z_n) \f_n
  e^{i\k\cdot(\r_m-\r_n)},
\end{align*} 
with
\begin{align*}
  \tQ(\k,z) :=& \int_\R \left( \frac{\id}{\nkf^2} - \frac{\kf \otimes
      \kf}{\nkf^4}\right) e^{i\kappa z}\d\kappa\\
  =& \int_\R \left( \frac{\id}{\nk^2+\kappa^2} - \frac{\kf \otimes
      \kf}{(\nk^2+\kappa^2)^2}\right) e^{i\kappa z}\d\kappa\\
  & \hspace{-11pt} =: \tQ_1 + \tQ_2,
\end{align*}
where $\tQ_1$ and $\tQ_2$ denote integrals over the diagonal- and
outer product terms respectively. Note that, since $\k\neq 0$, $k^2>0$
and the integrals are nonsingular. We shall encounter integrals of the
form
\begin{align*}
  L^p_q(k,z) := \int_0^\infty \frac{\kappa^p \cos(\kappa
    z)}{(k^2+\kappa^2)^q} \d \kappa \quad \text{and} \quad M^p_q(k,z)
  := \int_0^\infty \frac{\kappa^p \sin(\kappa z)}{(k^2+\kappa^2)^q} \d
  \kappa,
\end{align*}
for which we have, with $X=L,M$, that 
\begin{align}
  X^p_{q+1} = -\frac{1}{2kq} \pd{X^p_q }{k}, \quad q\in \Z, q
  >0. \label{eq:boosting_q}
\end{align}
Evidently,
\begin{align}
  \tQ_1(\k,z) = 2 \id L^0_1(\nk,z).
\end{align} 
The integral here is known \cite[3.723(2), p. 424]{Zwillinger2007}, 
\begin{align*}
  L^0_1(k,z) = \frac{\pi  e^{-k |z|}}{2 k}, \quad k>0.
\end{align*}
To simplify matters in computing $\tQ_2$ -- reducing the number of
integrals to labor over -- we note that the integrand can be viewed as
an \emph{element-wise} multiplication of
\begin{align*}
  (\k,1)\otimes(\k,1)=
  \begin{bmatrix}
    k_1^2 & k_1 k_2 & k_1 \\
    k_1 k_2 & k_2^2 & k_2 \\
    k_1 & k_2 & 1
  \end{bmatrix}
\end{align*}
and
\begin{align*}
  \frac{(1,1,\kappa)\otimes(1,1,\kappa)}{(\nk^2 + \kappa^2)^2} =
  (\nk^2 + \kappa^2)^{-2}
  \begin{bmatrix}
    1 & 1 & \kappa  \\
    1 & 1 & \kappa  \\
    \kappa & \kappa & \kappa ^2
  \end{bmatrix}
\end{align*}
where the former is a constant under the integration. The latter
demonstrates that there are three integrals to compute and arrange as:
\begin{align*}
  \tQ_2(\k,z) = -2
  \left[
    \begin{array}{rrr}
      k_1^2 L^0_2(\nk,z) & k_1k_2 L^0_2(\nk,z) & i k_1 M^1_2(\nk,z) \\
      k_1k_2 L^0_2(\nk,z) & k_2^2 L^0_2(\nk,z) & i k_2 M^1_2(\nk,z) \\
      i k_1 M^1_2(\nk,z) & i k_2 M^1_2(\nk,z) & L^2_2(\nk,z)
    \end{array}
  \right].
\end{align*}
Having already found $L^0_1$, \eqref{eq:boosting_q} gives that
\begin{align*}
  L^0_2(k,z) = \frac{\pi  e^{-k |z|} (k |z|+1)}{4 k^3}.
\end{align*}
Moreover, $L_2^2 = L_0^0 - k^2L^0_2$, so that
\begin{align*}
  L^2_2(k,z) = -\frac{\pi  e^{-k |z|} (k |z|-1)}{4 k},
\end{align*}
by \eqref{eq:boosting_q}. Finally,
\begin{align*}
  M^1_2(k,z) = -\pd{}{z} L^0_2(k,z) = \frac{\pi  z e^{-k |z|}}{4 k}.
\end{align*}
Thus, we have obtained an explicit form of the $\k$-space 2P Stokeslet
sum \eqref{eq:uf_deltas}: 
\begin{align*}
  \tU(\x_m) = \frac{4}{L^2} \sum_{\k\neq 0} \sumn \tQ(\k,z_m-z_n) \f_n
  e^{i\k\cdot(\r_m-\r_n)},
\end{align*}
where
\begin{align*}
  \tQ(\k,z) = 2 & \left[
    \begin{array}{ccc}
      L^0_{1}-k_1^2 L^0_{2} & -k_1 k_2 L^0_{2} & -i k_1 M^1_{2} \\
      -k_1 k_2 L^0_{2} & L^0_{1}-k_2^2 L^0_{2} & -i k_2 M^1_{2} \\
      -i k_1 M^1_{2} & -i k_2 M^1_{2} & L^0_{1}-L^2_{2}
    \end{array}
  \right] (\nk,z) \\
  \\
  =\frac{e^{-\nk |z| }}{\nk} &
  \begin{bmatrix}
    \pi -\frac{(\pi \nk |z|+\pi ) k_1^2}{2 \nk ^2} & -\frac{\pi (\nk
      |z|+1) k_1 k_2}{2 \nk ^2} &
    -\frac{1}{2} i \pi  z k_1 \\
    -\frac{\pi (\nk |z|+1) k_1 k_2}{2 \nk ^2} & \pi -\frac{(\pi \nk
      |z|+\pi ) k_2^2}{2 \nk ^2} &
    -\frac{1}{2} i \pi  z k_2 \\
    -\frac{1}{2} i \pi z k_1 & -\frac{1}{2} i \pi z k_2 & \frac{1}{2}
    (\pi \nk |z|+\pi )
  \end{bmatrix}.
\end{align*}

\subsection{Explicit $\k$-space form of regularized Stokeslet
  sum} \label{app:screened_u_integrals}

Our objective here is to compute the integral $Q$, given in
\eqref{eq:Q}, to obtain an explicit 2P Stokeslet Ewald sum. The tensor
$B$ \eqref{eq:hasimoto_fd} contains a diagonal and an outer product
term, and we integrate them separately. That is, let
\begin{align*}
  Q(\k,z) = \QI(\k,z) + \Qkk(\k,z),
\end{align*}
where 
\begin{align*}
  \QI(\k,z) &= \id\, e^{-\nk^2/4\xi^2} \int_\R \left( \frac{1}{4\xi^2}
    +
    \frac{1}{\nkf^2} \right) e^{-\kappa^2/4\xi^2} e^{i\kappa z}\d \kappa \\
  &= \id\,e^{-\nk^2/4\xi^2} \int_\R \left(\frac{1}{4\xi^2} +
    \frac{1}{\nk^2+\kappa^2} \right) e^{-\kappa^2/4\xi^2} e^{i\kappa
    z}\d \kappa
\end{align*}
and
\begin{align*}
  \Qkk(\k,z) &= - e^{-\nk^2/4\xi^2} \int_\R \left(
    \frac{1}{4\xi^2\nkf^2 } + \frac{1}{\nkf^4}\right) (\kf\otimes\kf)
  e^{-\kappa^2/4\xi^2}
  e^{i\kappa z}\d \kappa\\
  &= - e^{-\nk^2/4\xi^2} \int_\R \left( \frac{1}{4\xi^2(\nk^2 +
      \kappa^2)} + \frac{1}{(\nk^2+\kappa^2)^2} \right)
  (\kf\otimes\kf) e^{-\kappa^2/4\xi^2} e^{i\kappa z}\d \kappa.
\end{align*}
For non-negative integers $p$ and $q$ we let
\begin{align*}
  \tilde{J}^p_q(k,z) := \int_0^\infty \frac{\kappa^p \cos(\kappa
    z)}{(k^2+\kappa^2)^q} e^{-\kappa^2/4\xi^2} \d \kappa, \qquad
  J^p_q(k,z) := e^{-k^2/4\xi^2} \tilde{J}^p_q(k,z)
\end{align*}
and
\begin{align*}
  \tilde{K}^p_q(k,z) := \int_0^\infty \frac{\kappa^p \sin(\kappa
    z)}{(k^2+\kappa^2)^q} e^{-\kappa^2/4\xi^2} \d \kappa, \qquad
  K^p_q(k,z) := e^{-k^2/4\xi^2} \tilde{K}^p_q(k,z).
\end{align*}
With these definitions, it is evident that
\begin{align*}
  \QI(\k,z) = 2 \left( \frac{J^0_{0}(z)}{4\xi^2}  + J^1_{0}(\nk,z)\right) \id.
\end{align*}
The first integral is elementary,
\begin{align}
  \tilde{J}^0_0(z) = \sqrt{\pi } \xi  e^{-z^2 \xi ^2}, \label{eq:tJ00}
\end{align}
and the second can be found \cite[3.954(2), p. 504]{Zwillinger2007}
\begin{align}
  \tilde{J}^0_1(k,z) = \frac{\pi e^{k^2/4 \xi ^2} }{4k}\left(e^{-k z}
    \erfc\left(\frac{k}{2 \xi }-z \xi \right)+e^{k z}
    \erfc\left(\frac{k}{2 \xi }+z \xi \right)\right),\quad k>0. \label{eq:tJ01}
\end{align}
Moving on to $\Qkk$, one studies the symmetries of the integrand as
when computing $\tQ_2$, to find that
\begin{align*}
  \Qkk(\k,z) = -2\left[
    \begin{array}{ccc}
      k_1^2 \left(\frac{J^0_{1}}{4 \xi ^2}+J^0_{2}\right) & 
      k_1 k_2 \left(\frac{J^0_{1}}{4 \xi ^2}+J^0_{2}\right) & 
      k_1 \left(\frac{i K^1_{1}}{4 \xi ^2}+i K^1_{2}\right) \\
      k_1 k_2 \left(\frac{J^0_{1}}{4 \xi ^2}+J^0_{2}\right) & 
      k_2^2 \left(\frac{J^0_{1}}{4 \xi ^2}+J^0_{2}\right) & 
      k_2 \left(\frac{i K^1_{1}}{4 \xi ^2}+i K^1_{2}\right) \\
      k_1 \left(\frac{i K^1_{1}}{4 \xi ^2}+i K^1_{2}\right) & 
      k_2 \left(\frac{i K^1_{1}}{4 \xi ^2}+i K^1_{2}\right) & 
      \frac{J^2_{1}}{4 \xi ^2}+J^2_{2}
    \end{array}
  \right](\nk,z).
\end{align*}
The integrals present are related, as previously, via
\eqref{eq:boosting_q} and obvious algebraic relationships. With
$\tilde{J}^0_0$ and $\tilde{J}^0_1$ known, one finds
\begin{align*}
  \tilde{J}^0_2(k,z) = -\frac{1}{2k} \pd{}{k}\tilde{J}^0_1(k,z)\\
  \tilde{J}^2_1(k,z) = \tilde{J}^0_0(k,z) - k^2 \tilde{J}^0_2(k,z)\\
  \tilde{J}^2_2(k,z) = -\frac{1}{2k} \pd{}{k}\tilde{J}^2_1(k,z).
\end{align*}
Moving on to the anti-symmetric integrals $K^p_q$, one can find
\cite[3.954(1),p. 504]{Zwillinger2007}
\begin{align*}
  \tilde{K}^1_1(k,z) = \frac{\pi e^{k^2/4 \xi ^2}}{4} \left(e^{-k z}
    \erfc\left(\frac{k}{2 \xi }-z \xi \right)-e^{k z}
    \erfc\left(\frac{k}{2 \xi }+z \xi \right)\right),
\end{align*}
and, consequently,
\begin{align*}
  \tilde{K}^1_2(&k,z) = -\frac{1}{2k} \pd{}{k} \tilde{K}^1_1(k,z).
\end{align*}
This completes the delicate matter of integration. There remains to
find compact expressions for the $J^p_q$- and $K^p_q$ terms that were
left as derivatives above. The forms of \eqref{eq:tJ00} and
\eqref{eq:tJ01} suggest that we look for terms
\begin{align*}
  \lambda  &:= e^{-k^2/4\xi^2 - \xi^2 z^2}\\
  \theta_+ &:= e^{kz} \erfc\left( \frac{k}{2\xi} + \xi z\right)\\
  \theta_- &:= e^{-kz} \erfc\left( \frac{k}{2\xi} - \xi z\right).
\end{align*} 
With this, explicit forms of $J^p_q$ and $K^p_q$, as given in
\eqref{eq:J00}-\eqref{eq:K12} follow.

\section{Conventions for Fourier transforms and series}
\label{app:ft_defn}
For compactness, we use a non-unitary form of the Fourier
transform. For a function, $f(\x)$, $\x\in\R^n$, which is periodic
with respect to $\Omega \subset \R^n$, we have the Fourier series
defined as
\begin{align*}
  f(\x) &= \sum_\k \hat{f}_\k e^{i\k\cdot\x}\\
  \hat{f}_\k &= \frac{1}{|\Omega|} \int_\Omega f(\x) e^{-i\k\cdot\x} \d x,
\end{align*}
where $\k\in\{ 2\pi \mathbf{n}/L: \mathbf{n}\in\Z^n \}$.  The
corresponding integral transform for functions that decay sufficiently
fast on $\R^n$ is then naturally
\begin{align*}
  f(\x) &= \frac{1}{(2\pi)^n}\int_{\R^n} \hat{f}(\bar{\kappa})
  e^{i\bar{\kappa}\cdot \x}\d\bar{\kappa}\\
  \hat{f}(\bar{\kappa}) &= \int_{\R^n} f(\x) e^{-i \bar{\kappa}\cdot
    \x}\d \x, \quad \bar{\kappa} \in \R^n.
\end{align*}

\bibliographystyle{plain}
\bibliography{se2p_stokes}{}
\end{document}